\documentclass[12pt]{ectjstyle}
\usepackage{amsmath}
\usepackage{graphicx,psfrag,epsf}
\usepackage{enumerate}
\usepackage{adjustbox}
\usepackage{natbib}
\usepackage{url} % not crucial - just used below for the URL 
\usepackage{color}
\usepackage[flushleft]{threeparttable}
\usepackage[utf8]{inputenc}
\usepackage{subfig}
\usepackage{rotating}

\usepackage{bbm}

\usepackage{ifpdf}

\def \argmin{\mathop{\hbox{\rm arg min}}}

\newcommand{\bgamma}{\boldsymbol{\gamma}}
\newcommand{\balpha}{\boldsymbol{\alpha}}
\newcommand{\bbeta}{\boldsymbol{\beta}}

\newcommand{\btheta}{\boldsymbol{\theta}}

\newcommand{\bSigma}{\boldsymbol{\Sigma}}

\newcommand{\bI}{\boldsymbol{I}}

\newcommand{\bX}{\boldsymbol{X}}

\newcommand{\bZ}{\boldsymbol{Z}}

\usepackage{amsmath}
\usepackage{amssymb}
\usepackage{mathabx}
\usepackage{algorithm}
\usepackage{algorithmic}

\selectfont{}

\title[\,]{Estimation in high-dimensional linear regression: Post-Double-Autometrics as an alternative to Post-Double-Lasso}
       \author[\,]{Sullivan Hué$^{\dagger}$, 
       S\'{e}bastien Laurent$^{\dagger\ddagger}$, Ulrich Aiounou$^{\dagger}$ and Emmanuel Flachaire$^{\dagger}$\footnote{Corresponding author: Emmanuel Flachaire, Aix-Marseille University, AMSE, 5 bd Maurice Bourdet, 13001 Marseille, France, emmanuel.flachaire@univ-amu.fr}}
       \address{$^{\dagger}$Aix-Marseille University, CNRS, AMSE, France}
       \address{$^{\ddagger}$Aix-Marseille Graduate School of Management -- IAE and Institut Universitaire de France (IUF), France.}
\begin{document}

%\def\spacingset#1{\renewcommand{\baselinestretch}%
%{#1}\small\normalsize} \spacingset{1}

%%%%%%%%%%%%%%%%%%%%%%%%%%%%%%%%%%%%%%%%%%%%%%%%%%%%%%%%%%%%%%%%%%%%%%%%%%%%%%

%
%\if0\blind
%{
%  \title{\bf Estimation in high-dimensional linear regression: Post-Double-Autometrics as an alternative to Post-Double-Lasso\footnote{We thank participants at the ANR MLEforRisk workshop in Orléans, 2024 French Conference in Applied Econometrics using Stata, 2024 Quantitative Finance and Financial Econometrics conference in Marseille,  FinEML conference 2024 in Lugano and Autometrics workshop in Aix-en-Provence for their comments.We acknowledge research support by the French National Research Agency Grants ANR-17-EURE-0020 and ANR-21-CE26-0007-01 (project MLEforRisk) and by the Excellence Initiative of Aix-Marseille University - A*MIDEX.}}
%\author{Sullivan Hué\footnote{Aix-Marseille University (Aix-Marseille School of Economics), CNRS \& EHESS.}, S\'{e}bastien Laurent\footnotemark[2]  \footnote{Aix-Marseille Graduate School of Management -- IAE and Institut universitaire de France (IUF), France.}, Ulrich Aiounou\footnotemark[2] \hspace{0.02cm}  \& Emmanuel Flachaire\footnotemark[2]  \\}
%  \maketitle
%} \fi
%
%\if1\blind
%{
%  \bigskip
%  \bigskip
%  \bigskip
%  \begin{center}
%    {\LARGE\bf Estimation in high-dimensional linear regression: Post-Double-Autometrics as an alternative to Post-Double-Lasso}
%\end{center}
%  \medskip
%} \fi

\bigskip
\begin{abstract}
Post-Double-Lasso is becoming the most popular method for estimating linear regression models with many covariates when the purpose is to obtain an accurate estimate of a parameter of interest, such as an  average treatment effect. However, this method can suffer from substantial omitted variable bias in finite sample. We propose a new method called Post-Double-Autometrics, which is based on Autometrics, and show that this method outperforms Post-Double-Lasso. Its use in a standard application of economic growth sheds new light on the hypothesis of convergence from poor to rich economies.

\bigskip 
\noindent {\bf Keywords} Treatment effect, High dimension, Lasso, Autometrics.

\end{abstract}

%\noindent%
%{\it Keywords:}  Treatment effect, High dimension, Lasso, Autometrics.
%
%
%\noindent%
%{\it JEL codes:}  C21, C52, C55

%\newpage
%\spacingset{1.8} % DON'T change the spacing!

\section{Introduction}
\label{sec:intro}

%Post-Lasso \citep{belloni2013} and 

In high-dimensional linear regression,  covariate selection becomes a key issue when the purpose is to obtain a good estimate of a parameter of interest,  such as an average treatment effect.  It is important not to miss  relevant variables correlated to the outcome and the treatment variable (to avoid bias),  and not to include irrelevant variables (to avoid loss of accuracy).
For this purpose, Post-Double-Lasso (\citealp{belloni2014}) is becoming the most popular method. This method allows the use of many covariates, to avoid omitted variable bias, while controlling overfitting issue and loss of accuracy with a Lasso variable selection method.

%Post-Double Lasso \citep{belloni2014} is becoming the most popular method for estimating a parameter of interest, as an average treatment effect,  in high-dimensional linear regression. This method allows the use of many covariates, to avoid omitted variable bias, while controlling overfitting issue and loss of accuracy with a Lasso variable selection method.

However, \citet{wuthrich2023} demonstrate that this method can experience significant omitted variable bias in finite samples, even in straightforward settings favorable to Lasso. They also show that the performance of this method is very sensitive to the choice of the regularization parameter of the Lasso, and no specific choice of this parameter systematically outperforms the others. %{\color{red} @@} They do not, however, propose a solution to mitigate these issues. {\color{red} j'enleverai cette phrase qui a une connotation critique ("they do not") envers un potentiel referee}

In this paper, we consider an alternative variable selection method, Autometrics (\citealp{doornik2009}), which is based on statistical inference. This method performs automatic selection of variables based on ``Hendry's theory of reduction'' (see Chapter 6 of \citealp{Hendry2014}) and an automatic general-to-specific model selection (\citealp{Hendry2000}). Its main hyperparameter, the target size $\alpha$, is the significance level used for inference. Simulation results show that this parameter allows to determine approximatively the frequency of irrelevant explanatory variables included in the terminal model (\citealp{Hendry2014}, \citealp{epprecht2021}, \citealp{flachaire2024}).

\begin{figure}
	\centering
	\includegraphics[width=.9\linewidth, height=.3\textheight]{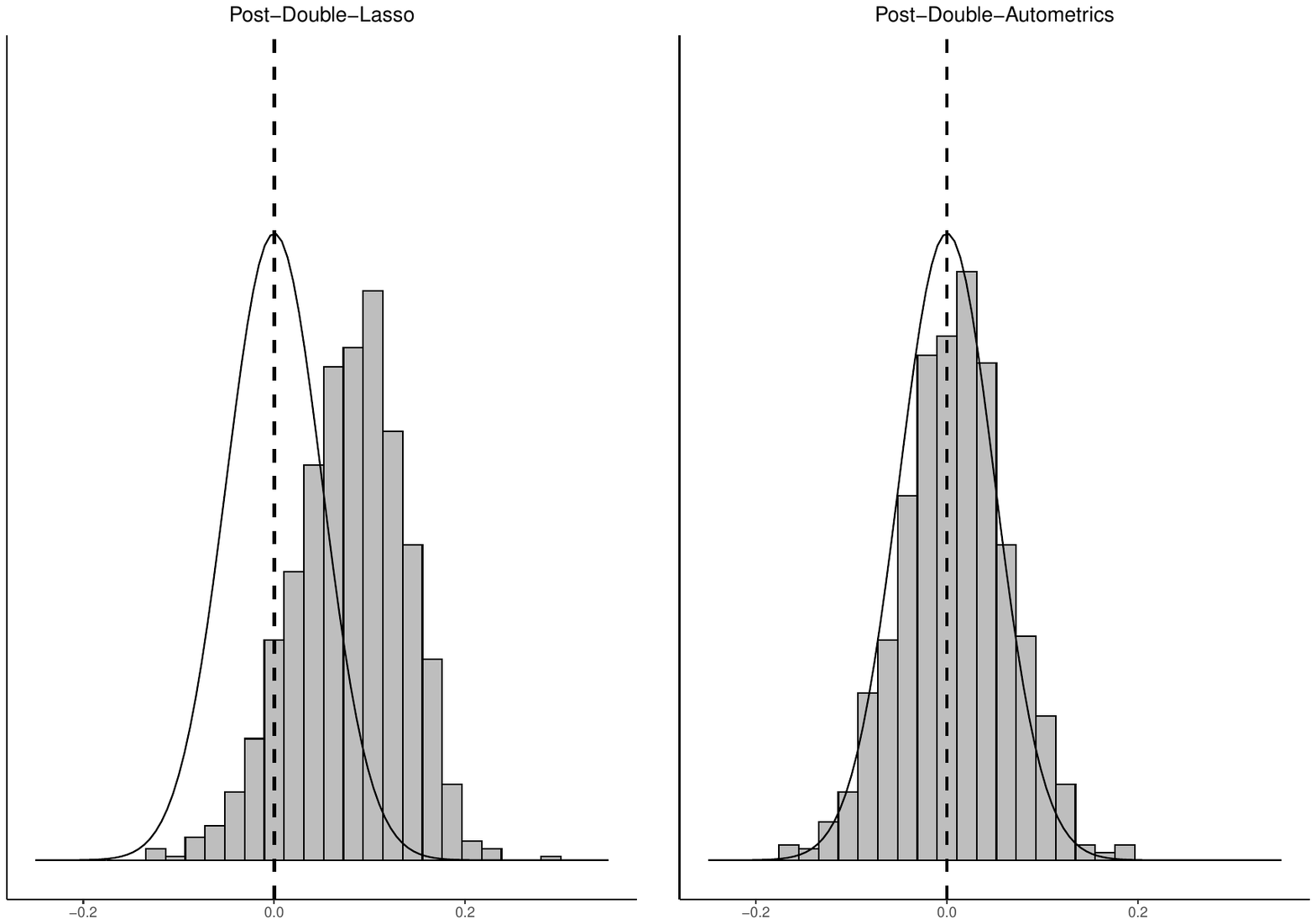}
	\caption{Empirical distribution of the Post-Double-Lasso (left panel) and Post-Double-Autometrics  (right panel) estimation of the treatment effect.  In both panels, the distribution of the oracle estimator corresponds to the solid line while the true value corresponds to the dotted vertical line.}
	\label{fig:intro_histogram}
\end{figure}

Firstly, we propose a new method called Post-Double-Autometrics, that extends Post-Double-Lasso by using Autometrics instead of Lasso to do variable selection. 
We show through extensive Monte Carlo experiments that the proposed method provides the best variable selection method and outperforms Post-Double-Lasso in many cases. To illustrate, Figure \ref{fig:intro_histogram} provides a numerical illustration of this phenomenon in a simple simulation experiment, with $n=400$ observations and $p=210$ covariates of which only $10$ are relevant.  The error term is {\em i.i.d.} and the covariates are moderately correlated. The Lasso penalization parameter $\lambda^{min}$ is obtained by cross-validation, and the Autometrics target size  is 5\%.
%(with a correlation ranging between $-0.6$ and $0.36$). 
The left panel shows the behavior of the Post-Double-Lasso estimator  of a treatment effect whose coefficient is equal to $0$
while the right panel shows the empirical distribution of our proposed Post-Double-Autometrics estimator.
In this setting, the Post-Double-Lasso estimator is upward biased while the Post-Double-Autometrics estimator  is approximately unbiased and behaves almost like the oracle estimator. 
More details on this simulation will be given  in Section \ref{sec_corr}. %in Section \ref{section_simulations} and more particularly in Section \ref{sec_corr}.
%The left panel shows the behavior of the Post-Double-Lasso estimator (with penalization parameter $\lambda$ set to $\lambda^{min}$ as discussed in the next section) of a treatment effect whose coefficient is equal to $0$ while the right panel shows the empirical distribution of our proposed Post-Double-Autometrics estimator (with a target size of 5\%). 
%In this setting, the Post-Double-Lasso estimator is upward biased while the Post-Double-Autometrics estimator  is approximately unbiased and behaves almost like the oracle estimator. More details on this simulation will be given in Section \ref{section_simulations} and more particularly in Section \ref{sec_corr}.
%(with penalization parameter $\lambda$ set to $\lambda^{min}$ as discussed in the next section)  (with a target size of 5\%)

%Overall, our  simulation results  show that Post-Double-Autometrics provides the best variable selection method and outperform Post-Double-Lasso when the latter does not perform well. 

Secondly, we revisit a standard economic growth application, on the convergence hypothesis that the poor catch up the rich countries, with the \cite{Barro1994} data consisting of a panel of countries for the period of 1960 to 1985. This hypothesis is often tested from a linear regression of growth rates on the initial level of GDP and additional characteristics of countries. Given that the number of covariates that describe such characteristics can be comparable to the sample size, covariate selection is an issue. Empirical results based on Lasso selection method support the convergence hypothesis (\citealp{Belloni2011b}, \citealp{Belloni2013c}). We show that this conclusion is sensitive to the choice of the penalization term, and that the convergence hypothesis is rejected with Post-Double-Autometrics whatever the target size. Further analysis shows that including or missing one specific covariate is most likely the source of support or lack of support for the convergence hypothesis.

%Autometrics is known to have very good properties in the sense that, unlike Lasso, its main hyperparameter (i.e., the target size $\alpha$) allows to determine approximatively the frequency of irrelevant explanatory variables included in the terminal model {\color{red} (@@ references)}. Importantly, like the hyperparameter of Lasso (i.e., $\lambda$), the target size also influences the frequency of relevant variables selected in the terminal model. However, we show in this paper that there are cases where Autometrics fails to keep some relevant variables, which translates into a  biased estimate of the average treatment effect when these variables are correlated with the treatment variable. 

%To overcome this problem, we propose a new method called Post-Double-Autometrics, that extends Post-Double-Lasso by using Autometrics instead of Lasso to do variable selection. 
%We show through Monte Carlo experiments and an empirical illustration that the proposed method outperforms Post-Double-Lasso in some situations. Figure \ref{fig:intro_histogram} provides a numerical illustration of this phenomenon in a simple simulation experiment, with $n=500$ observations and $p=200$ covariates. The Post-Double-Lasso estimator of the treatment effect is biased, shifted much to the right compared to the true value. To the opposite, the Post-Double-Autometrics estimator is approximately unbiased and behaves almost like the oracle estimator.
%Our  simulation results also show that Post-Double-Autometrics provides the best variable selection method. 

The reminder of the paper is structured as follows. Section 2 presents the pitfalls of Lasso-based methods and introduce our new approach. Section 3 is devoted to Monte Carlo experiments and Section 4 to an empirical application. The paper concludes with Section 5.

%%%%%%%%%%%%%%%%%%%%%%%%%%%%%%%%%%%%%%%%%%%%%%%%%%%%%%%%%%%%%%%%%%%%%%%
\section{Estimation in high-dimensional linear regression} \label{Methods_section}

%In this section, we first show the pitfalls of Post-Lasso and Post-Double-Lasso, then introduce Post-Double-Autometrics.

In this section, we first present the framework, Post-Double-Lasso and finally Post-Double-Autometrics.

%\subsection{Pitfalls of Post-Lasso and Post-Double-Lasso}
\subsection{Framework}

In accordance with the literature, we consider the following standard linear regression model \begin{align}
	y_i &= \bX_i \bbeta + d_i \delta + \epsilon_i,  \label{eqY} \\
	d_i &= \bX_i \bgamma + \eta_i, \label{eqD} 
\end{align}
for $ i=1,\dots,n$, where $y_i$ is the outcome, $d_i$ is  % a dummy variable corresponding to 
the treatment variable, $\bX_i = \left(x_{1,i}, \dots, x_{p,i}\right) \in \mathbb{R}^p$ is a vector of $p$ covariates, and $\epsilon_i$ and $\eta_i$ are error terms. The parameter of interest ${\delta}$ is the average treatment effect.
The covariates might include transformations of raw variables in order to capture non-linear effects (e.g., powers of raw variables, interaction terms). %However, unlike Double Machine Learning \citep{Chernozhukov2018}, we assume that these effects are introduced parametrically in Equations \eqref{eqY}-\eqref{eqD}. 

In observational studies,  the unconfoundedness condition is crucial to obtain  unbiased estimation of the average treatment effect  ${\delta}$. This condition requires  $\bX$ to contain all the confounder variables which explain both the outcome $y$ and the treatment  $d$. Otherwise, the estimation may suffer from substantial omitted variable bias (OVB) or counfounding bias.
In practice, it is difficult to account for all counfounder variables. Here, we consider the strategy of including  many covariates and  using a variable selection method to select the relevant covariates (to make the estimate as efficient as possible). 

\subsection{Post-Double-Lasso}

An early approach developed in the literature is Lasso, which performs estimation and variable selection simultaneously  (\citealp{tibshirani1996}). Considering the linear regression in (\ref{eqY}), Lasso  solves the following penalized regression problem
\begin{equation}
\hat{\btheta} = \underset{\bbeta}{\argmin}\left[\sum_{i=1}^{n}\left(y_i - \bZ_i \btheta\right)^2 + \lambda \sum_{j=1}^{p}\omega_j|\beta_j|\right],
\label{eq:lasso}
\end{equation}
where $\bZ_i=(\bX_i,d_i)$, $\btheta=(\bbeta^\top,\delta)^\top$ and $\omega_j = 1 \ \forall j$. Estimation is feasible in a high dimensional setting, that is, when the number of covariates $p$ is large and can exceed the number of observations $n$. The penalization restricts the magnitude of the coefficients, making some of them exactly zero with a sufficiently large $\lambda$. Several approaches have been proposed in the literature to set $\lambda$. 
For example, \citet{bickel2009} proposed to set $\lambda^{bya} = 2\sigma \sqrt{2n^{-1}\left(1+\tau\right)\log p}$, where $\sigma$ is the standard deviation of the error term, and $\tau>0$.
\cite{belloni2012} proposed a penalization choice, denoted as $\lambda^{bcch}$, to take into account heteroscedastic and clustered errors.
%\cite{belloni2012} proposed a penalization choice, denoted as $\lambda^{bcch}$, to take into account heteroscedastic and clustered errors, and which is recommended by \citet{belloni2014} for Post-Double-Lasso.
Another popular approach is to set $\lambda$ by cross-validation, with the value minimizing the prediction error, ${\lambda}^{min}$, or the value associated to the most parsimonious model within a 1-standard-error interval, ${\lambda}^{1se}$. See \citet{Homrighausen2013,Homrighausen2014} and \citet{Chetverikov2021} for theoretical results on the choice of $\lambda$ by cross-validation.

The penalization shrinks all  coefficients towards zero, introducing some bias into the coefficient estimates associated with  relevant covariates. Since the interest is on the estimation of the parameter $\delta$, it is recommended to use the Post-Lasso (\citealp{belloni2013}) which relies on two steps. First, estimate \eqref{eqY} by Lasso to select a subset of variables, denoted  $\bX^*$. Second, estimate $\delta$ from the OLS regression of $y$ on $\bX^*$ and $d$. 
To mitigate the bias introduced in the coefficients of relevant variables, the Adaptive-Lasso can also be used (\citealp{Zou2006}). It solves the penalized regression problem in  (\ref{eq:lasso}),  with weights $\omega_j = |\hat{\mu}_j|^{-\eta}$, where $\eta > 0$ and $\hat{\mu}_j$ is a consistent estimator of $\theta_j$. Here, smaller weights are assigned to larger coefficients in the constraint and, thus, large non-zero coefficients shrink more slowly to zero as $\lambda$ increases.

%Note that the Lasso can be replaced by one of its extensions in the first step. For example, \eqref{eq:lasso} corresponds to the adaptive-Lasso \citep{Zou2006} when $\omega_j = |\hat{\mu}_j|^{-\eta}$, where $\eta > 0$ and $\hat{\mu}_j$ is a consistent estimator of $\theta_j$.

%The penalization shrinks all penalized coefficients towards zero, introducing some bias in the estimates. Since the interest is on the estimation of the parameter $\delta$, it is recommended to use the Post-Lasso  \citep{belloni2013} described below:
%\begin{enumerate}
%\item use Lasso  on (\ref{eqY}) to select a subset of  variables, denoted  $\bX^*$;
%\item estimate $\delta$ from the OLS regression of $y$ on $d$ and $\bX^*$.
%\end{enumerate}

\cite{belloni2014} show that Lasso usually fails to select some relevant variables in the first step, which may lead to a biased estimate of the parameter of interest $\delta$. This typically occurs when some covariates in $\bX$ are correlated with $d$, but only have a small effect on $y$. To solve this issue, they proposed the Post-Double-Lasso. The standard linear regression model defined in \eqref{eqY}-\eqref{eqD} implies the following reduced form model:
\begin{equation}
	y_i = \bX_i\balpha + \nu_i,
\end{equation}
where $\balpha = \bbeta+\bgamma \delta$ and $\nu_i = \epsilon_i + \eta_i\delta$. In order to select relevant control variables, Post-Double-Lasso relies on the following two steps of Lasso 
\begin{align}
	\hat{\balpha} = \underset{\balpha}{\argmin}\left[\sum_{i=1}^{n}\left(y_i - \bX_i\balpha\right)^2 + \lambda_1 \sum_{j=1}^{p}{w}_{1,j}|\alpha_j|\right], \\
	\hat{\bgamma} = \underset{\bgamma}{\argmin}\left[\sum_{i=1}^{n}\left(d_i - \bX_i\bgamma\right)^2 + \lambda_2 \sum_{j=1}^{p}{w}_{2,j}|\gamma_j|\right], 
\end{align}
where ${w}_{1,j} = {w}_{2,j} = 1 \ \forall j$. Finally, $\delta$ is estimated from the OLS regression of $y$ on $d$ and the union of selected control variables, such as
\begin{equation}
	\left(\hat{\bbeta},\hat{\delta}\right) = \underset{\bbeta,\delta}{\argmin}\left[\sum_{i=1}^{N}\left(y_i  - \bX_i\bbeta -d_i\delta \right)^2\right] \ s.t. \ \beta_j = 0 \ \forall j \notin \{\hat{\bI}_{\bX^{*}} \cup \hat{\bI}_{\bX^{**}}\},
\end{equation}
where $\hat{\bI}_{\bX^{*}} = supp\left(\hat{\balpha}\right) = \{j:\hat{\alpha}_j \ne 0\}$ and $\hat{\bI}_{\bX^{**}} = supp\left(\hat{\bgamma}\right) = \{j:\hat{\gamma}_j \ne 0\}$.
This method thus gives covariates omitted in the first step a chance to be selected in the second step, especially those correlated to $d$ and likely to lead to an omitted variable bias. Similarly to the Post-Lasso, several choices can be considered for the penalization parameters $\lambda_1$ and $\lambda_2$, and the Lasso can be replaced by the Adaptive-Lasso. %one its extension, including the Adaptive-Lasso.%  whith different weights ${w}_{1,j}$ and ${w}_{2,j}$.% are different from 1 $\forall j$. %Note that in the following $\lambda_1$ and $\lambda_2$ are set according to the same approach, and are referred to as $\lambda$ for the sake of clarity.

%\cite{belloni2014} show that Lasso may fail to select some relevant variables in the first step, which may lead to a biased estimate of the parameter of interest $\delta$. This typically occurs when some control variables in $\bX$ are correlated with $d$, but only have a small effect on $y$. To solve this issue, they proposed the Post-Double-Lasso:
%\begin{enumerate}
%\item use Lasso on a restricted version of (\ref{eqY}) in which $d$ is ignored to select a subset of  variables  $\bX^*$;
%\item use Lasso on  (\ref{eqD})  to select a subset of  variables  $\bX^{**}$;
%\item estimate $\delta$ from the OLS regression of $y$ on $d$ and the union of $\bX^*$ and $\bX^{**}$.
%\end{enumerate}
%This method gives control variables omitted in the first step a second chance to be recovered in the second step, especially those correlated to $d$ and likely to lead to an omitted variable bias.

Although the Post-Double-Lasso is a substantial improvement over the Post-Lasso, \citet{wuthrich2023} show that it can still lead to an OVB. 
%\textcolor{red}{More precisely, under some assumptions, the authors show theoretically that the probability to not select any relevant control variable, given that there is at least one, goes to 1 if $p$ is relatively large. This result occurs when the magnitude of the coefficients of control variables is not large enough compared to the penalty parameter. In this case, the Lasso misses relevant variables in both steps of Post-Double-Lasso. However, coefficients of control variables are still large enough to cause a bias in the estimation of $\delta$. Moreover, they also show through simulation experiments that the extent of the bias depends on the choice of the regularization parameter, and that there is no single choice of $\lambda$ among popular ones systematically leading to better results. }
Specifically, under certain assumptions, the authors theoretically demonstrate that the probability of not selecting any relevant control variables, given that at least one exists, becomes high when $p$ is relatively large. This occurs when the magnitude of the control variable coefficients is not sufficiently large compared to the penalty parameter. In such cases, the Lasso method may fail to capture relevant variables in both steps of the Post-Double-Lasso procedure and induce an OVB on $\delta$. 
%\textcolor{blue}{The authors provide no reliable solution to this issue.}
Additionally, the authors show through simulations that the magnitude of the bias depends on the choice of the regularization parameters $\lambda_1$ and $\lambda_2$, and that no single popular choice  consistently leads to better results. 
%They do not, however, propose a solution to mitigate these issues.

%Although the Post-Double-Lasso is a substantial improvement over the Post-Lasso, \citet{wuthrich2023} show that it can still lead to  OVB issue, when the coefficients of relevant variables are not large enough for the Lasso to select them but large enough to cause bias in the estimation of ${\delta}$. Moreover, they show that the extent of the bias depends on the choice of the regularization parameter, and that there is no single choice of $\lambda$ among popular ones systematically leading to better results. 

\subsection{Post-Double-Autometrics} %: An inference-based method }

To address this OVB issue, we propose to use another variable selection method:  Autometrics (\citealp{doornik2009}). 
Autometrics is a popular variable selection model among econometricians
because it relies on statistical inference to select the relevant variables.
There exists at least six different software implementations of automatic ``general-to-specific'' (GETS) model selection (Matlab, OxMetrics, Scilab, STATA, EViews and R). Autometrics is one of them and is implemented in OxMetrics. See \citet{Sucarrat2020} for more details on general-to-specific modelling and its implementations.

More specifically, Autometrics is an algorithm performing automatic variables selection  with the GETS model selection approach 
(\citealp{hendry1999}, \citealp{Hendry2000}). As Lasso, this method starts from a generalized unrestricted model (GUM) which includes all potential relationships between the outcome $y$ and the control variables $\bX$, such as lag values, trends, step indicators or any other transformations of the raw variables aiming at capturing non-linearities. The algorithm then performs a battery of tests to remove insignificant variables from $\bX$ and to find a congruent parsimonious model. The final model is then estimated by OLS.\footnote{For more details on Autometrics, see \citet{doornik2009} and \citet{Castle_etal2023}.}
Using a multiple block-search algorithm, Autometrics can also be used with more observations than covariates.

Autometrics has three main advantages. First, unlike the Lasso performing variable selection  based on the magnitude of coefficients (\citealp{wuthrich2023}), Autometrics relies on statistical inference. Therefore, Autometrics-based methods will select variables whenever their coefficients are significant, independently of the magnitude of their coefficients. Second, the tuning parameter of Autometrics is the targeted nominal size $\alpha$.  It allows the user to control the expected proportion of irrelevant variables included in the final model. This approach contrasts sharply with Lasso, where $\lambda$ directly controls the number (or proportion) of variables selected in the final model. %\footnote{Typical values of this parameter are $\alpha=0.05$, or $\alpha=0.01$ for a more conservative user.} 
Third, Autometrics may invalidate a reduction when the reduced model does not pass some pre-defined misspecification tests.

The target size $\alpha$ refers to the pre-specified significance level or threshold that guides the inclusion or exclusion of variables in the model during the selection process of Autometrics. Specifically, it determines the acceptable probability of incorrectly rejecting a true null hypothesis (i.e., including a variable that is not truly relevant). By setting a target size, Autometrics controls the false-positive rate of variable selection, balancing between overfitting (selecting too many irrelevant variables) and underfitting (failing to select relevant variables).
In practical terms:
\begin{itemize}
\item A smaller target size (e.g., $\alpha = 0.01$) is more conservative, leading to fewer variables being selected as only those with strong statistical significance are retained.
\item A larger target size (e.g., $\alpha = 0.05$) is less stringent, allowing more variables to be included, even if their significance is somewhat weaker.
\end{itemize}
The choice of target size directly impacts the number of control variables selected, which can influence both the results and the robustness of the model. 
%\textcolor{blue}{@@ Rajouter un mot dur le multiple testing ?@@}
However, unlike for Lasso, the tuning parameter of Autometrics is more straightforward and intuitive to select. 

Lasso methods and Autometrics both start with a general model containing potentially many covariates, from which the goal is to select the most relevant ones. However, unlike Lasso, which estimates parameters through a single optimization process, Autometrics employs an iterative procedure, reducing the model using statistical tests and a tree search algorithm. As a result, deriving theoretical results for Autometrics is more challenging, and its properties are primarily assessed through Monte Carlo simulations, as shown in \citet{Hendry2011}. Nonetheless, theoretical insights on similar general-to-specific model selection procedures and algorithms can be found in \citet{Johansen2009}, \citet{Hendry2011}, \citet{Hendry2015}, and \citet{Johansen2016}.

Autometrics allows to force the presence of variable $d$ in Equation \eqref{eqY} in the terminal model. However, Autometrics is expected to miss relevant control variables with a low degree of non-centrality (i.e., true $t$-statistic) and, if these variables are strongly correlated with $d$, this can lead to an OVB for $\delta$, similarly to Post-Lasso.

%Autometrics allows to force the presence of variable $d$ in Equation \eqref{eqY} in the terminal model. However, Autometrics is expected to miss relevant control variables with a low degree of non-centrality (i.e., true $t$-statistic) and, if these variables are strongly correlated with $d$, this can lead to an OVB for $\delta$.

Therefore, following the approach of \citet{belloni2014}, we propose an extension of Autometrics, which we call Post-Double-Autometrics. Our method is similar to the Post-Double-Lasso described above, with Lasso  being replaced by Autometrics in the first two steps. By giving covariates omitted in the first step a chance to be recovered in the second step, this method is expected to improve over the usual implementation of Autometrics. We expect Post-Double-Autometrics to address the OVB issue of Post-Double-Lasso, as Autometrics selects variables based on statistical inference rather than coefficient magnitude. This means Autometrics will not overlook relevant variables, even if their coefficient magnitudes are small, as long as they are statistically significant.

Post-Double-Autometrics shares similarities with the indicator saturation method proposed by \citet{hendry2006}. Their approach consists in splitting the initial sample in two blocks, and applying a general-to-specific variable selection method on each block. A final variable selection is then applied on the union of variables selected in each block. However, the dependent variable considered in both blocks of the indicator saturation method is the same ($y$). Moreover, the main objective of the indicator saturation method is to detect outliers or breaks and not estimating an average treatment effect as in the Post-Double-Autometrics.

%%%%%%%%%%%%%%%%%%%%%%%%%%%%%%%%%%%%%%%%%%%%%%%%%%%%%%%%%%%%%%%%%%%%%%%%%%%%%
\section{Simulations}
\label{section_simulations}

In this section, we use Monte Carlo simulations to study the finite sample properties of different methods for estimating a parameter of interest in high-dimensional linear regression. We consider two  single-selection methods, the naive Post-Lasso and Autometrics, and three double-selection methods, the benchmark Post-Double-Lasso,  the extension Post-Double-Adaptive-Lasso, and our proposed Post-Double-Autometrics.\footnote{For the adaptive-Lasso, we set $\eta = 1$ and $\hat{\mu}_j$ to the estimate of an OLS regression when $p+1 < N/2$ and a ridge regression otherwise.}  Monte Carlo simulations are performed on $1{,}000$ replications.
%The Monte Carlo simulation is inspired by \citet{wuthrich2023} and is carried out on $1,000$ replications. 

The data generating process (DGP) corresponds to Equations \eqref{eqY}-\eqref{eqD}, where $\bX_i \overset{i.i.d}{\sim} \mathcal{N}\left(\mathbf{0_p}, \bSigma_{\bX}\right)$, $\bSigma_{\bX}$ is a Toeplitz matrix with element $(k,l)$ set to $\rho^{|k-l|}$ for $k,l=1,\ldots,p$, $\epsilon_i \overset{i.i.d}{\sim} \mathcal{N}\left(0, \sigma^2_\epsilon\right)$ and $\eta_i \overset{i.i.d}{\sim} \mathcal{N}\left(0, \sigma^2_\eta\right)$. %, with $\epsilon_i~\bot~\eta_i$.
The parameter $\rho$ controls the degree of correlation between the covariates. The DGP is general enough to  consider cases where variables are positively or negatively correlated (when $\rho$ is respectively positive or negative) but also weakly or strongly correlated (i.e., when $|\rho|$ is respectively close to 0 or to 1). 

To assess the variable selection properties of the different methods, we consider a few relevant variables ($\beta_j\neq0,\gamma_j\neq0$, for $j=1,\dots,10$) and many irrelevant variables ($\beta_j=\gamma_j=0$, for $j=11,\dots,p$). Instead of choosing arbitrary values for the non-zero coefficients, we calibrate their individual statistical significance as follows:
\begin{align}
	\beta_j &= \psi_j^y \sqrt{\sigma^2_\epsilon \left[\bSigma_{\bZ}^{-1}\right]_{jj}}, \\
	\gamma_j &= \psi_j^d \sqrt{\sigma^2_\eta \left[\bSigma_{\bX}^{-1}\right]_{jj}},
\end{align}
where $\psi_j^y$ and $\psi_j^d$ are the non-centrality parameters of the relevant covariates $\bX_j$, $j=1,\dots,10$, in \eqref{eqY}-\eqref{eqD}, $\bZ = \left(\bX, d\right)$, and 
\begin{equation*}
	\bSigma_Z = 
	\begin{pmatrix}
		\bSigma_{\bX} & \bSigma_{\bX}\bgamma  \\
		\bgamma'\bSigma_{\bX}  & \bgamma'\bSigma_{\bX} \bgamma + \sigma^2_{\eta}
	\end{pmatrix}.
\end{equation*}
The values of  $\psi_j^y$ and $\psi_j^d$ correspond to the expected $t$-statistics for, respectively, the null hypotheses $H_0: \beta_j=0$ in  Equation \eqref{eqY} and  $H_0: \gamma_j=0$ in Equation  \eqref{eqD}. They allow to control the relevance of covariates, since the probability to reject a null hypothesis is high (low) for  large (small) value of the  non-centrality parameter.
To simplify the notation, we will denote by $\psi^y$ (resp. $\psi^d$)
the non-centrality parameters of all non-zero coefficients $\beta_j$ (resp. $\gamma_j$) of the 10 relevant covariates $\bX_j$, $j=1,\dots,10$.

To study the finite sample performance of the different methods mentioned above, we report the average bias and root mean squared error (RMSE) of $\hat\delta$ as well as the proportion of selected relevant (potency) and irrelevant (gauge) variables included in the terminal model. High potency is important to avoid biased estimation, while for a given level of potency, a smaller gauge is expected to lead to more efficiency (i.e., lower variance of $\hat\delta$).  

\subsection{Independent covariates}

In the first simulation, we set $n=400$, $p=210$, $\rho = 0$, i.e., $\bSigma_X=I_p$, $\sigma^2_\epsilon = \sigma^2_\eta = 1$, and our parameter of interest $\delta =0$. In this case, the covariates are orthogonal which is the most favorable case for all methods. 
Moreover, we use $\psi_j^y = 2.5$ and $\psi_j^d = 4$  for $j = 1, \dots, 10$, so that  coefficients of the relevant variables have expected $t$-statistics equal to $2.5$ for  $H_0: \beta_j=0$ in  Equation \eqref{eqY}, and equal to 4 for $H_0: \gamma_j=0$ in Equation  \eqref{eqD}. The probability for these variables to be kept in the final model is therefore quite low in the first equation, and higher in the second equation. In this framework, double-selection methods are expected to outperform single-selection methods.

Table \ref{tab:rho0} shows bias and RMSE of the treatment effect $\hat\delta$, as well as the average proportion of relevant (potency) and irrelevant (gauge) variables.
Note that for the  Post-Double-Lasso and Post-Double-Autometrics the gauge and potency are computed on the union of $\bX^*$ and $\bX^{**}$, and we consider several choices of tuning parameters $\lambda$ and $\alpha$. 
%Values in bold  are those corresponding to  Figure \ref{fig:intro_histogram}. 
%Several comment can be made here.

First, double-selection methods provide substantial improvements over single selection methods, regardless of the choice of the variable selection method considered. The potency is always increased in the double-selection approaches, leading to lower bias and RMSE. %Such results are expected as the probability to select relevant variables is small in Equation \eqref{eqY}, but close to 1 in Equation \eqref{eqD}. 
Overall, the results  suggest that single-selection methods are always biased in this setting, unlike double-selection methods. Therefore, we focus on double-selection methods in the following and leave single-selection methods out of the discussion.

\begin{table}[tb]
	\centering
	\caption{Bias, RMSE , Gauge and Potency of  $\hat\delta$ with independent covariates ($\rho = 0$). Design: $\psi^y = 2.5, \psi^d = 4$, $n=400$, $p=210$.}
	\renewcommand{\arraystretch}{1.0}
\resizebox{13cm}{!}{	
	\begin{tabular}{lcccc}
		\hline
		\multicolumn{1}{l}{Model \,\, ($\psi^y = 2.5, \psi^d = 4$)} & \multicolumn{1}{l}{Bias} & \multicolumn{1}{l}{RMSE} & \multicolumn{1}{l}{Potency} & \multicolumn{1}{l}{Gauge} \\
		\hline
		OLS   & 0.0010 & 0.0749 & 1.0000  & 1.0000 \\
		\hline
		Post-Lasso, $\lambda^{min}$ & 0.1036 & 0.1298 & 0.3868 & 0.0532 \\
		Post-Lasso, $\lambda^{1se}$ & 0.1723 & 0.1802 & 0.0350 & 0.0024 \\
		Post-Lasso, $\lambda^{bya}$ & 0.1803 & 0.1856 & 0.0000 & 0.0000 \\
		Post-Lasso, $\lambda^{bcch}$ & 0.0833 & 0.1032 & 0.4857 & 0.0487 \\
		\hline
		Post-Double-Lasso, $\lambda^{min}$ & 0.0011 & 0.0598 & 0.9999 & 0.2156 \\
		Post-Double-Lasso, $\lambda^{1se}$ & 0.0343 & 0.0698 & 0.7718 & 0.0192 \\
		Post-Double-Lasso, $\lambda^{bya}$ & {0.1660} & {0.1720} & {0.1000} & {0.0000} \\
		Post-Double-Lasso, $\lambda^{bcch}$ & 0.1690 & 0.1758 & 0.0454 & 0.0000 \\
		\hline
		Post-Double-Adaptive-Lasso, $\lambda^{min}$ & 0.0017 & 0.0597 & 1.0000 & 0.3212 \\
		Post-Double-Adaptive-Lasso, $\lambda^{1se}$ & 0.0070 & 0.0564 & 0.9479 & 0.0858 \\
		\hline
		%Autometrics, $\alpha = 0.10$ & 0.0623 & 0.0976 & 0.5420 & 0.0761 \\
		Autometrics, $\alpha = 0.05$ & 0.0832 & 0.1139 & 0.4342 & 0.0364 \\
		Autometrics, $\alpha = 0.01$ & 0.1253 & 0.1451 & 0.2316 & 0.0088 \\
		\hline
		Post-Double-Autometrics, $\alpha = 0.05$ & {0.0097} & {0.0556} & {0.9056} & {0.0821} \\
		Post-Double-Autometrics, $\alpha = 0.01$ & 0.0264 & 0.0603 & 0.8059 & 0.0204 \\
		\hline
	\end{tabular}
}
%	\caption{Bias and RMSE of  $\hat\delta$, proportion of relevant (potency) and irrelevant (gauge) selected variables, with independent covariates ($\rho = 0$). Design: $\psi^y = 2.5, \psi^d = 4$, $n=400$, $p=210$.}
	\label{tab:rho0}
\end{table}

Second, the performance of Post-Double-Lasso methods depends on the choice of the $\lambda$ penalty term. Most of the relevant variables are missed (potency close to 0) with $\lambda^{bya}$ and $\lambda^{bcch}$, leading to a treatment effect poorly estimated (large bias and RMSE). 
The bias is close to 0 for Post-Double-Lasso with $\lambda^{min}$ as almost all relevant variables are selected (potency close to 1). However, $\lambda^{min}$ selects too many (more than 20\%, i.e. more than 40) irrelevant variables which translates into a small loss of efficiency. %Note that OLS corresponds to either Post-Lasso or Post-Double-Lasso with $\lambda=0$ and it is also found to be unbiased but much less efficient than Post-Double-Lasso with $\lambda^{min}$. 
Interestingly, Post-Double-Lasso with $\lambda^{1se}$ selects very few irrelevant variables (gauge close to 2\%) but also misses more than 20\% of relevant variables which translated into a much larger bias.
%there is an over-selection (higher gauge) with the latter, leading to a slight loss of accuracy (small increase of the RMSE). %, unlike Post-Double-Lasso with $\lambda^{bya}$. 
%These results come from the fact that Post-Double-Lasso is more parsimonious with some penalty terms than others. While Post-Double-Lasso is extremely parsimonious with $\lambda^{bya}$ and select very few variables, it is the opposite with $\lambda^{min}$ and it selects many control variables, both relevant and irrelevant. Therefore, Post-Double-Lasso is biased with $\lambda^{bya}$ as it does not include enough variables in the terminal model, whereas it is unbiased with $\lambda^{min}$ but at the cost of a relatively large gauge because it includes too many variables. Results obtained for $\lambda^{1se}$ and $\lambda^{bcch}$ are similar to those of $\lambda^{bya}$. 
Results suggest that $\lambda^{min}$ is the best choice for the penalty term of Post-Double-Lasso methods as it leads to an unbiased estimate of $\delta$ and a small RMSE.

Third, both the Post-Double-Adaptive-Lasso  with $\lambda^{min}$ and $\lambda^{1se}$ are unbiased and have a RMSE slightly lower than Post-Double-Lasso with $\lambda^{min}$. Post-Double-Adaptive-Lasso  with  $\lambda^{1se}$ performs particularly well because it has a high potency (close to 95\%) and a gauge lower than 10\%.
%slightly improves the performance standard Post-Double-Lasso methods. The bias and RMSE are reduced for Post-Double-Adaptive-Lasso with $\lambda^{1se}$, due to a larger potency. For $\lambda^{min}$, the results are similar. However, the use of the adaptive-lasso over the lasso leads to a small increase of the gauge.

Fourth, Post-Double-Autometrics leads to large potency, and thus to small bias, regardless of the choice of $\alpha$. Moreover, as Autometrics controls the proportion of irrelevant selected variables, while selecting most relevant variables, the gauge of Post-Double-Autometrics remains relatively small unlike Post-Double-(Adaptive-)Lasso with $\lambda^{min}$. The Post-Double-Autometrics method applies Autometrics successively to two different models and the expected gauge of each model is $\alpha$. Therefore, if the number of redundant variables is very large (like in our simulation), approximatively $2 \alpha$\% of redundant variables are expected to be retained when considering the union between $\bX^*$ and $\bX^{**}$, which is exactly what we observe in Table \ref{tab:rho0}.\footnote{Note that the gauge for the standard Autometrics method is approximately equal to $\alpha$.} 

Finally, although OLS using all variables is unbiased, its RMSE is relatively large compared to those of unbiased double-selection methods. For example, the RMSE of OLS is about $35\%$ larger than Post-Double-Autometrics with $\alpha = 0.05$. %and therefore, as expected, a good double-selection method provide more accurate estimates of the treatment effect than OLS.

To conclude, Post-Double-Autometrics with $\alpha=0.05$ competes very well with Post-Double-Lasso  with $\lambda^{min}$ and Post-Double-Adaptive-Lasso with $\lambda^{min}$ and $\lambda^{1se}$ in this very simple and unrealistic framework of independent covariates.

\subsection{Correlation and relevance of  covariates\label{sec_corr}}
%\subsection{Additional simulation results: effect of control covariates correlation and relevance}

This section complements the previous simulation results by first allowing the covariates to be correlated and then by varying the non-centrality parameter $\psi^d$ of the 10 relevant covariates in Equation \eqref{eqD}.
%\begin{figure}[htbp]
%	\centering
%	\includegraphics[width=\linewidth, height=.4\textheight]{./Graphs/n_rep_1000_n_400_p_10_k_200_psiYX_2.5_psiYD_0_psiDX_4_s2eps_1_s2eta_1_4_Graphs_new_1step.pdf} 
%	\caption{Bias, RMSE, Gauge and Potency of $\hat\delta$, with varying dependent control variables $\rho \in  [-0.9 ; 0.9]$. Design: $\psi^y = 2.5$, $\psi^d = 4$, $n=400$, $p=210$.}
%	\label{fig:simus_rho_single_selection_methods} %Code Get_results_5.ox
%\end{figure}

\begin{figure}[tbp]
	\centering
	\includegraphics[width=\linewidth, height=.4\textheight]{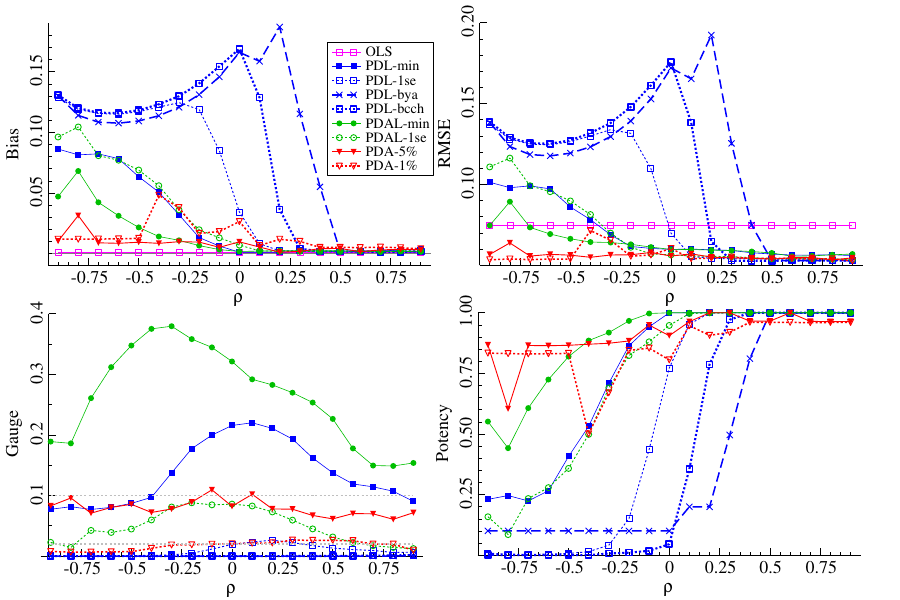} 
	\caption{Bias, RMSE, Gauge and Potency of $\hat\delta$ with dependent covariates $\rho \in  [-0.9 ; 0.9]$. Design: $\psi^y = 2.5$, $\psi^d = 4$, $n=400$, $p=210$.}
%	\caption{Bias, RMSE, Gauge and Potency of $\hat\delta$ as a function of $\rho$ for $\rho \in  [-0.9 ; 0.9]$. Design: $\psi^y = 2.5$, $\psi^d = 4$, $n=400$, $p=210$.}
	\label{fig:simus_rho_double_selection_methods} %Code Get_results_5.ox
\end{figure}

%\marginpar{SL: je dois ajouter OLS a la Figure 2}
Figure \ref{fig:simus_rho_double_selection_methods} displays bias and RMSE of $\hat\delta$ as well as the potency and gauge for $\psi^y = 2.5$,  $\psi^d = 4$, $n=400$ and $p=210$ dependent control variables, with $\rho \in \left[-0.9; 0.9\right]$, for OLS and double-selection methods, across $1{,}000$ replications.  The case $\rho=0$ corresponds to the results in Table \ref{tab:rho0}.\footnote{Results for single-selection methods are available upon request.}

The conclusions from Table \ref{tab:rho0} are confirmed, but some interesting new results are emerging. Indeed, Post-Double-Lasso methods are again found to be sensitive to the selection method of $\lambda$, Post-Double-Lasso with $\lambda^{min}$ can lead to more accurate estimates than OLS when $\rho>-0.4$, and the Post-Double-Adaptive-Lasso improves the results compared to standard Lasso methods. Importantly, the performance of all Lasso-based methods deteriorates when $\rho$ becomes strongly negative, i.e., the bias and RMSE increase while the potency largely decreases.\footnote{\citet{Hendry2014} found similar results, demonstrating that Lasso's effectiveness decreases significantly when variables are negatively correlated.}
%While most lasso-based methods provide satisfactory results for $\rho > 0$, and even outstanding ones when $\rho > 0.5$ (bias, RMSE and gauge close to 0, potency close to 1), their performances deteriorates substantially for $\rho < 0$. Indeed, the bias/RMSE (resp. potency) of lasso-based methods increase (resp. decrease) as $\rho$ decreases towards -0.9, regardless of the selection method of $\lambda$ and of the use of adaptive-lasso or not.

In contrast, Post-Double-Autometrics provides reliable results regardless of $\alpha$ and $\rho$. The bias and RMSE of Post-Double-Autometrics remain close to 0 for all values of $\rho$, whereas the gauge is still close to $2\alpha$. The results thus suggest that Post-Double-Autometrics outperforms other estimation methods as it provides the best treatment effect estimation (lowest bias and RMSE) and the best variable selection (high potency and controlled gauge), while being insensitive to   $\alpha$ and to the magnitude of the correlation between the covariates.

\begin{figure}[tbp]
	\centering
	\includegraphics[width=\linewidth, height=.4\textheight]{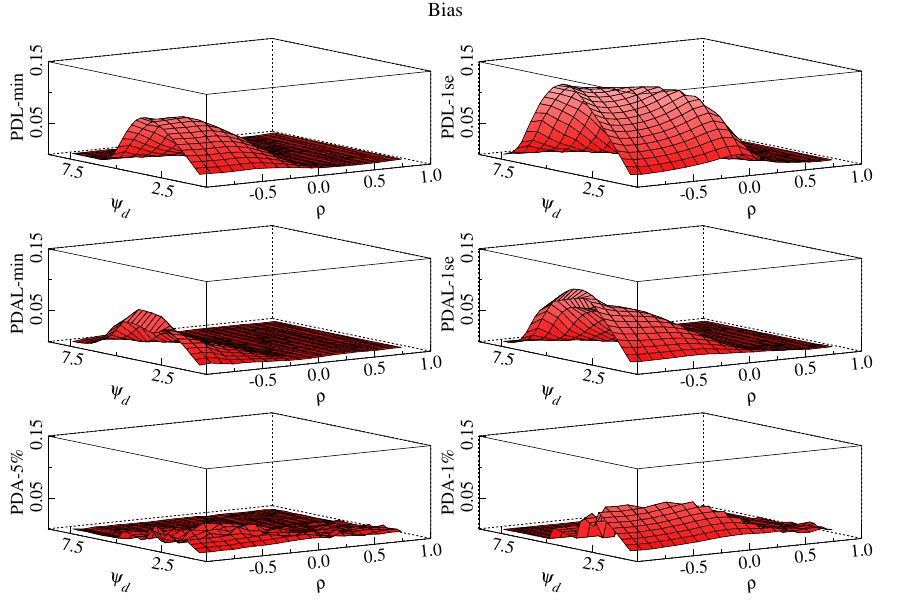} 
	\caption{Bias of $\hat\delta$, with dependent covariates $\rho \in  [-0.9 ; 0.9]$ and varying non-centrality measure $\psi^d \in [1 ; 8]$. Design: $\psi^y = 2.5$, $n=400$, $p=210$.}
%	\caption{Bias of $\hat\delta$, with varying non-centrality measure $\psi^d \in [1 ; 8]$ and dependent covariates $\rho \in  [-0.9 ; 0.9]$. Design: $\psi^y = 2.5$, $n=400$, $p=210$.}
%	\includegraphics[width=\linewidth, height=.4\textheight]{./Graphs/Bias_n_rep_1000_n_400_p_10_k_200_psiYD_0_psiDX_4_s2eps_1_s2eta_1.pdf} 
%	\caption{Bias of $\delta$, with varying non-centrality measure $\psi^y \in [1 ; 8]$ and dependent control variables $\rho \in  [-0.9 ; 0.9]$. Design: $\psi^d = 4$, $n=400$, $p=210$.}
	\label{fig:simus_3D_bias} %Get_results_all_PsiYX_2
\end{figure}

\begin{figure}[htbp]
	\centering
	\includegraphics[width=\linewidth, height=.4\textheight]{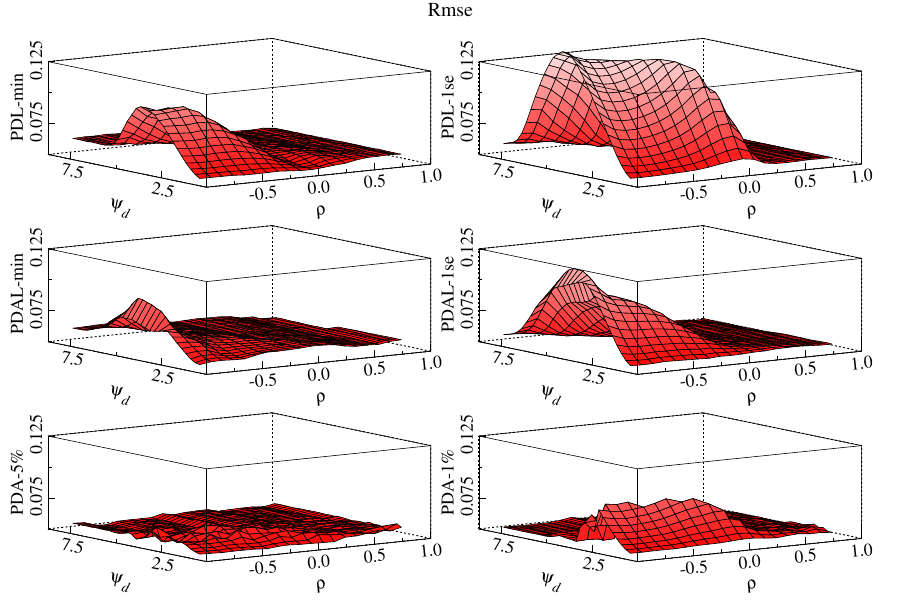} 
	\caption{RMSE of $\hat\delta$, with dependent covariates $\rho \in  [-0.9 ; 0.9]$ and varying non-centrality measure $\psi^d \in [1 ; 8]$. Design: $\psi^y = 2.5$, $n=400$, $p=210$.}
%	\caption{Rmse of $\hat\delta$, with varying non-centrality measure $\psi^d \in [1 ; 8]$ and dependent covariates $\rho \in  [-0.9 ; 0.9]$. Design: $\psi^y = 2.5$, $n=400$, $p=210$.}
%	\includegraphics[width=\linewidth, height=.4\textheight]{./Graphs/Rmse_n_rep_1000_n_400_p_10_k_200_psiYD_0_psiDX_4_s2eps_1_s2eta_1.pdf} 
%	\caption{RMSE of $\delta$, with varying non-centrality measure $\psi^y \in [1 ; 8]$ and dependent control variables $\rho \in  [-0.9 ; 0.9]$. Design: $\psi^d = 4$, $n=400$, $p=210$.}
	\label{fig:simus_3D_rmse} %Get_results_all_PsiYX_2
\end{figure}

\begin{figure}[htbp]
	\centering
	\includegraphics[width=\linewidth, height=.4\textheight]{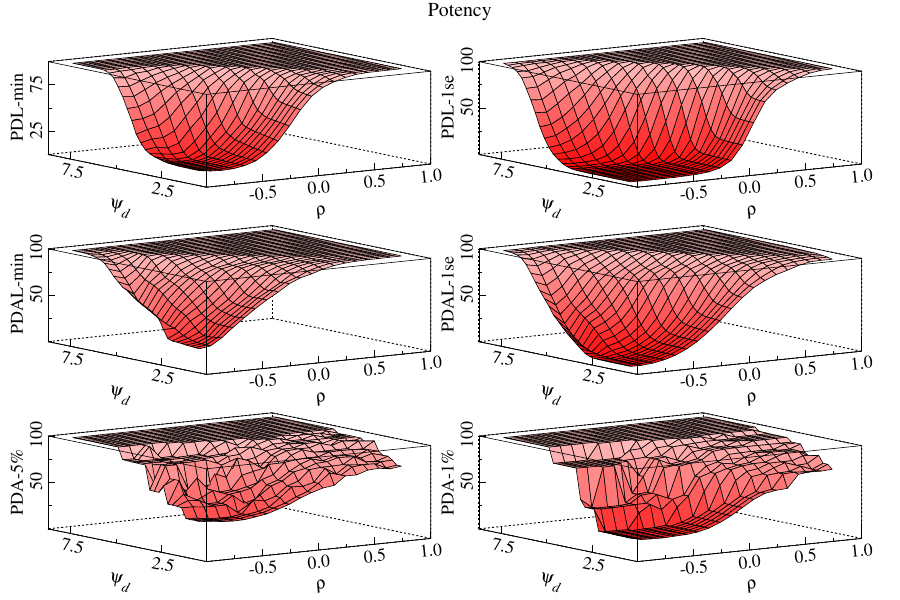} 
	\caption{Potency of $\hat\delta$, with dependent covariates $\rho \in  [-0.9 ; 0.9]$ and varying non-centrality measure $\psi^d \in [1 ; 8]$. Design: $\psi^y = 2.5$, $n=400$, $p=210$.}
%	\caption{Potency of $\hat\delta$, with varying non-centrality measure $\psi^d \in [1 ; 8]$ and dependent covariates $\rho \in  [-0.9 ; 0.9]$. Design: $\psi^y = 2.5$, $n=400$, $p=210$.}
%	\includegraphics[width=\linewidth, height=.4\textheight]{./Graphs/Potency_n_rep_1000_n_400_p_10_k_200_psiYD_0_psiDX_4_s2eps_1_s2eta_1.pdf} 
%	\caption{Potency of $\delta$, with varying non-centrality measure $\psi^y \in [1 ; 8]$ and dependent control variables $\rho \in  [-0.9 ; 0.9]$. Design: $\psi^d = 4$, $n=400$, $p=210$.}
	\label{fig:simus_3D_potency} %Get_results_all_PsiYX_2
\end{figure}

\begin{figure}[htbp]
	\centering
	\includegraphics[width=\linewidth, height=.4\textheight]{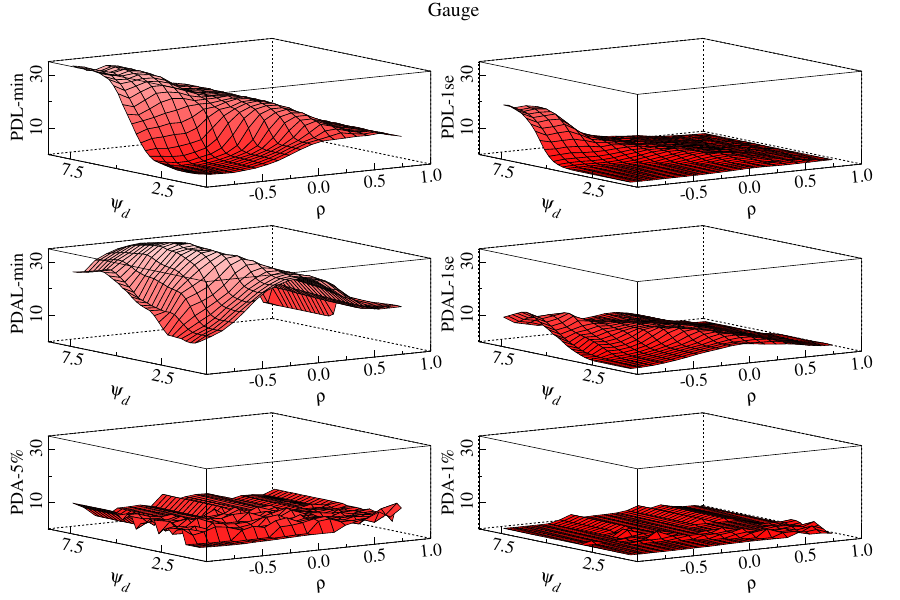} 
	\caption{Gauge of $\hat\delta$, with dependent covariates $\rho \in  [-0.9 ; 0.9]$ and varying non-centrality measure $\psi^d \in [1 ; 8]$. Design: $\psi^y = 2.5$, $n=400$, $p=210$.}
%	\caption{Gauge of $\hat\delta$, with varying non-centrality measure $\psi^d \in [1 ; 8]$ and dependent covariates $\rho \in  [-0.9 ; 0.9]$. Design: $\psi^y = 2.5$, $n=400$, $p=210$.}
%	\includegraphics[width=\linewidth, height=.4\textheight]{./Graphs/Gauge_n_rep_1000_n_400_p_10_k_200_psiYD_0_psiDX_4_s2eps_1_s2eta_1.pdf} 
%	\caption{Gauge of $\delta$, with varying non-centrality measure $\psi^y \in [1 ; 8]$ and dependent control variables $\rho \in  [-0.9 ; 0.9]$. Design: $\psi^d = 4$, $n=400$, $p=210$.}
	\label{fig:simus_3D_gauge} %Get_results_all_PsiYX_2
\end{figure}

Finally, Figures \ref{fig:simus_3D_bias}-\ref{fig:simus_3D_gauge} complement previous results by displaying respectively the bias, RMSE, potency and gauge as a function of the correlation coefficient, with $\rho\in [-0.9,0.9]$, and of the non-centrality parameter $\psi^d$, with $\psi^d \in [1,8]$, for $\psi^y = 2.5$,  $n=400$ and $p=210$. The higher $\psi^d$ is, the greater the probability that these variables will be selected, i.e., the more relevant they are. Therefore, the potency is expected to increase with $\psi^d$, resulting in lower values of the bias and RMSE. 
%The case $\psi^d = 4$ corresponds to the results displayed in Figure \ref{fig:simus_rho_double_selection_methods}.\footnote{Note that increasing $\psi^y$ or $\psi^d$ leads to similar results for double-selection methods.} 
Note that the case $\rho = -0.6$ and $\psi^d = 4$ corresponds to the results displayed in Figure \ref{fig:intro_histogram}, while those of Figure \ref{fig:simus_rho_double_selection_methods} have been obtained for $\psi^d = 4$ and $\rho\in [-0.9,0.9]$. %\footnote{Note that increasing $\psi^y$ or $\psi^d$ leads to similar results for double-selection methods.} 
For the sake of clarity, we focus on the best performing methods, i.e., Post-Double-Lasso for $\lambda = \{\lambda^{min}, \lambda^{1se}\}$, Post-Double-Adaptive-Lasso for $\lambda = \{\lambda^{min}, \lambda^{1se}\}$  and Post-Double-Autometrics for $\alpha = \{5\%, 1\%\}$.\footnote{Results for other methods are available upon request.} In the figures, these methods are denoted, respectively, \emph{PDL-min}, \emph{PDL-1se}, \emph{PDAL-min}, \emph{PDAL-1se}, \emph{PDA-5\%} and \emph{PDA-1\%}.

%The bias of lasso-based methods does not only depend on $\lambda$ and $\rho$, but also on $\psi_y$. Indeed, these methods remain biased even for large values of $\psi_y$ as the bias only disappears  from $\psi_y = 6$. Such results come from the potency that remains relativelly small until $\psi_y = 6$. Qualitatively similar results are observed for the RMSE. Moreover, results also suggest that the gauge increases with $\psi_y$, and even reaches levels above $30\%$ with $\lambda^{min}$. Similar results are observed with the adaptive-lasso, even though the results are slightly better (lower bias and RMSE). To summarize, double-selection methods based on lasso can lead to accurate treatment effect estimate, but their results depend on the choice of $\lambda$, the correlation within control covariates ($\rho$) and their individual relevance ($\psi_y$).

The bias in Lasso-based methods is influenced not only by the parameters $\lambda$ and $\rho$, but also by the value of $\psi^d$. Even for large $\psi^d$, these methods remain biased when $\rho$ is strongly negative, and the bias only disappears when $\psi^d \ge 7.5$. Similar patterns are observed for the RMSE. Furthermore, the results indicate that the gauge increases with $\psi^d$, reaching levels above 30\% with $\lambda^{min}$. Comparable findings hold for the adaptive-Lasso methods, though with slightly better performance in terms of bias and RMSE. In summary, while Lasso-based double-selection methods can yield accurate estimates of treatment effects, their performance depends on the choice of $\lambda$, the correlation among covariates ($\rho$), and the individual relevance of these covariates ($\psi^d$).

In contrast, the performance of Post-Double-Autometrics is remarkable in all cases. While the bias remains close to 0 for all values of $\psi^d$ due to a  potency close to 1, the gauge does not increase and is close to $2\alpha$ regardless of the value of $\psi^d$. Moreover, the RMSE of Post-Double-Autometrics is the lowest among all methods, and especially for $\alpha = 0.05$. Results thus suggest that Post-Double-Autometrics outperforms other methods as it provides the most accurate treatment effect, regardless of the choice of $\alpha$, the correlation between the covariates and the non-centrality parameter $\psi^d$. 

%\begin{enumerate}
%\item
%{\em Top left graph}: Post-Double-Autometrics  methods ({\tt PDA}) have a bias close to zero, while Post-Double-Lasso methods ({\tt PDL})  can have a large bias and they are very sensitive to the choice of $\lambda$ and $\rho$.
%\item
%{\em Top right graph}: Post-Double-Autometrics   ({\tt PDA}) are the most efficient methods, with smallest RMSE values close to 0.5. In contrast, Post-Double-Lasso ({\tt PDL}) is very sensitive to the method used to select $\lambda$ and to the value of $\rho$. 
%\item {\em Bottom graphs}: High potency is important to avoid biased estimation, while small gauge would provide more accuracy. Post-Double-Autometrics ({\tt PDA}) is the best variable selection method, with high potency and controlled gauge. 
%\end{enumerate}

%% Mettre Graph 3D pour Tout sauf bya et bcch
%% Robustesse pour n=200 uniquement Adaptive et PDA

%Y-axis: 0 8.5 2 2 1
%X-axis: -1 1 -0.9 0.4 0.2
% Copy axis, Labelling

%World coordinates
% Z1 0.05 Z2 0.15
%Y-axis: 0 8.5 2 2 1
%X-axis: -1 1 -0.9 0.4 0.2
%Z-axis: 0.05 0.15 0.05 0.05 0.025
% Copy World coordinates Z + axis, Labelling

\begin{figure}[h!]
	\centering
	\includegraphics[width=.48\linewidth, height=.40\textheight]{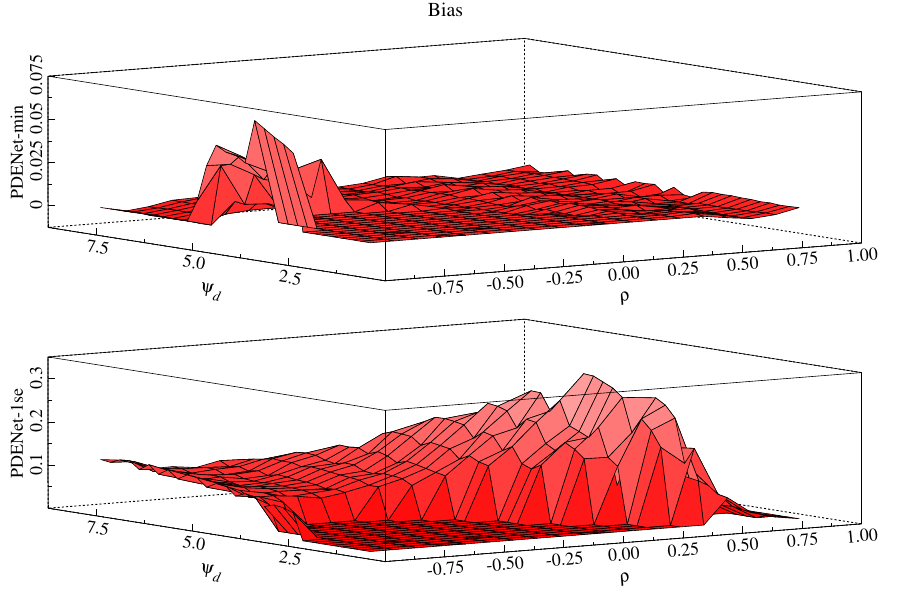} 
	\hfill
	\includegraphics[width=.48\linewidth, height=.40\textheight]{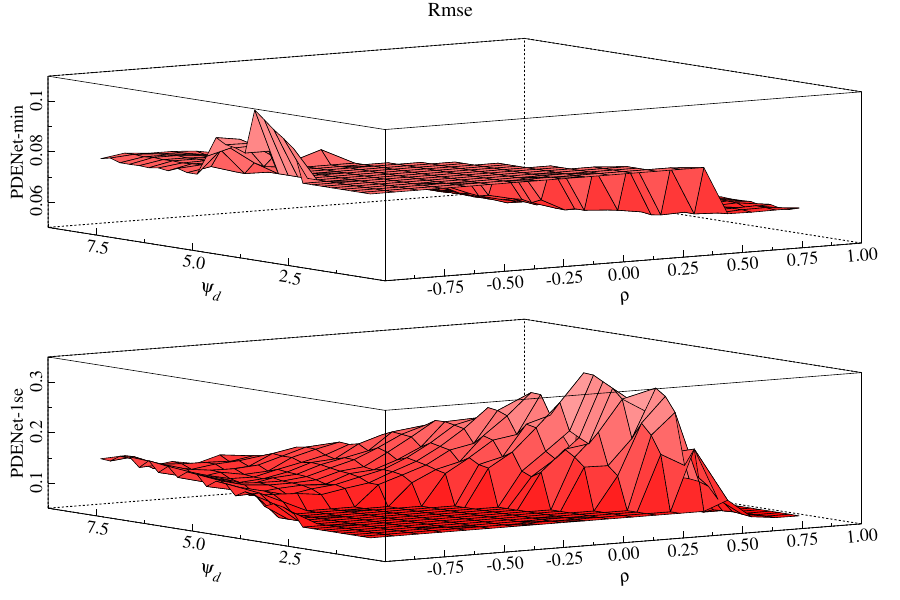} 
%	\caption{Post-Double-ElasticNet: Bias and RMSE of $\hat\delta$, with  $\rho \in  [-0.9 ; 0.9]$ and  $\psi^d \in [1 ; 8]$. Design: $\psi^y = 2.5$, $n=400$, $p=210$.}
%	\caption{Bias of $\hat\delta$, with varying non-centrality measure $\psi^d \in [1 ; 8]$ and dependent covariates $\rho \in  [-0.9 ; 0.9]$. Design: $\psi^y = 2.5$, $n=400$, $p=210$.}
%	\includegraphics[width=\linewidth, height=.4\textheight]{./Graphs/Bias_n_rep_1000_n_400_p_10_k_200_psiYD_0_psiDX_4_s2eps_1_s2eta_1.pdf} 
%	\caption{Bias of $\delta$, with varying non-centrality measure $\psi^y \in [1 ; 8]$ and dependent control variables $\rho \in  [-0.9 ; 0.9]$. Design: $\psi^d = 4$, $n=400$, $p=210$.}
%	\label{fig:simus_3D_bias_ENet} %Get_results_all_PsiYX_2
%\end{figure}
%\begin{figure}[hbp]
%\medskip

%\bigskip
	\centering
	\includegraphics[width=.48\linewidth, height=.40\textheight]{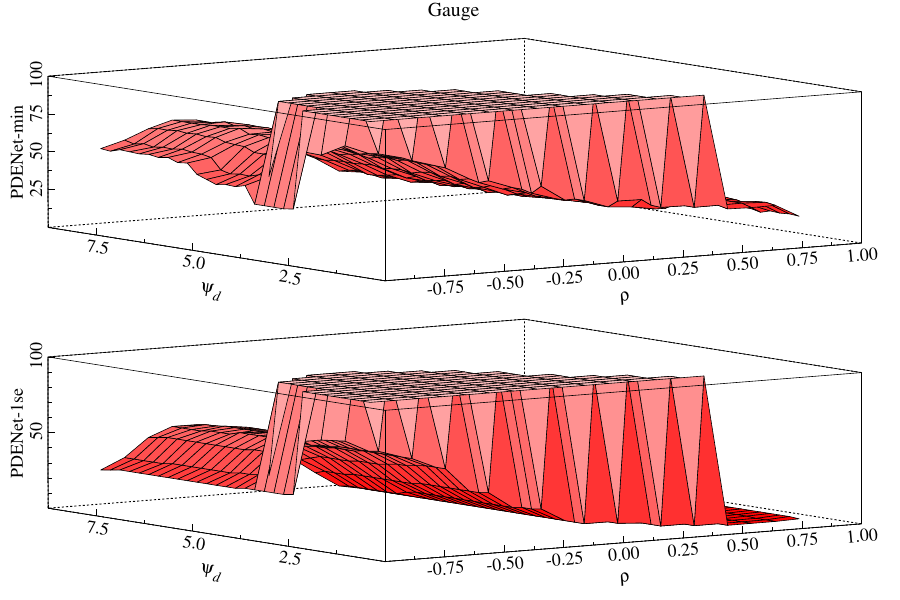} 
	\hfill
	\includegraphics[width=.48\linewidth, height=.40\textheight]{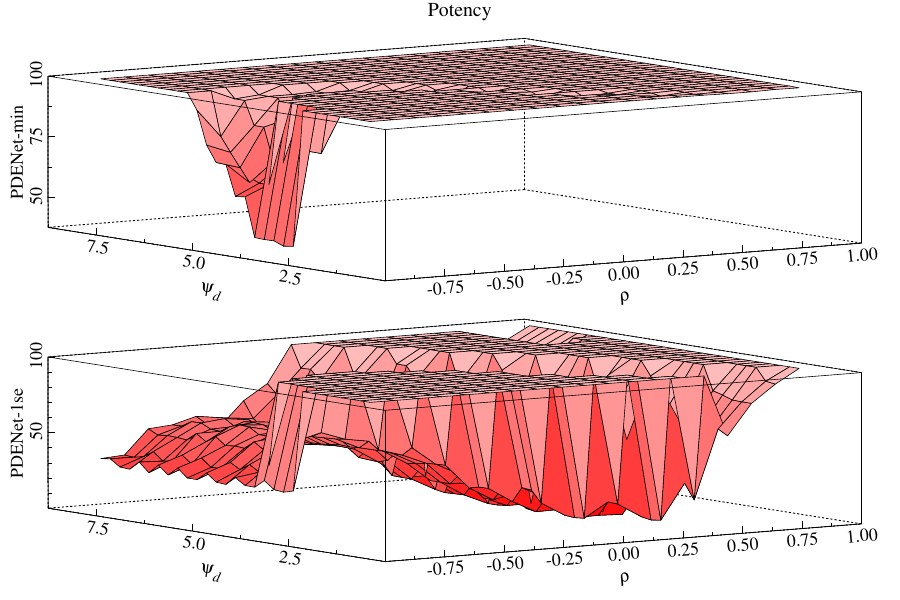} 
	\caption{Post-Double-ElasticNet: Bias, RMSE, Gauge and Potency of $\hat\delta$, with  dependent covariates $\rho \in  [-0.9 ; 0.9]$ and varying non-centrality measure  $\psi^d \in [1 ; 8]$. Design: $\psi^y = 2.5$, $n=400$, $p=210$ and $100$ simulations.}
%	\caption{Bias of $\hat\delta$, with varying non-centrality measure $\psi^d \in [1 ; 8]$ and dependent covariates $\rho \in  [-0.9 ; 0.9]$. Design: $\psi^y = 2.5$, $n=400$, $p=210$.}
%	\includegraphics[width=\linewidth, height=.4\textheight]{./Graphs/Bias_n_rep_1000_n_400_p_10_k_200_psiYD_0_psiDX_4_s2eps_1_s2eta_1.pdf} 
%	\caption{Bias of $\delta$, with varying non-centrality measure $\psi^y \in [1 ; 8]$ and dependent control variables $\rho \in  [-0.9 ; 0.9]$. Design: $\psi^d = 4$, $n=400$, $p=210$.}
	\label{fig:simus_3D_ENet} %Get_results_all_PsiYX_2
\end{figure}

In Appendix 2, additional simulation results are presented with Elastic Net, which combines Lasso and Ridge penalizations. This method can overcome some limitations of the Lasso, especially with group of highly correlated covariates (\citealp{zouhastie2005}). Used in a post-double selection procedure,  Figure \ref{fig:simus_3D_ENet} shows better results than Post-Double-Lasso (lower bias) but worse than  Post-Double-Autometrics (higher gauge and RMSE) in Figures \ref{fig:simus_3D_bias}-\ref{fig:simus_3D_gauge}. Note that the Elastic Net method requires the calibration of two hyperparameters, which greatly increases the calculation time, reason why we only carried out 100 replications for this method.

\subsection{More covariates than observations} % ($p>n$)}

\begin{figure}[tbp]
	\centering	
	\includegraphics[width=\linewidth, height=.4\textheight]{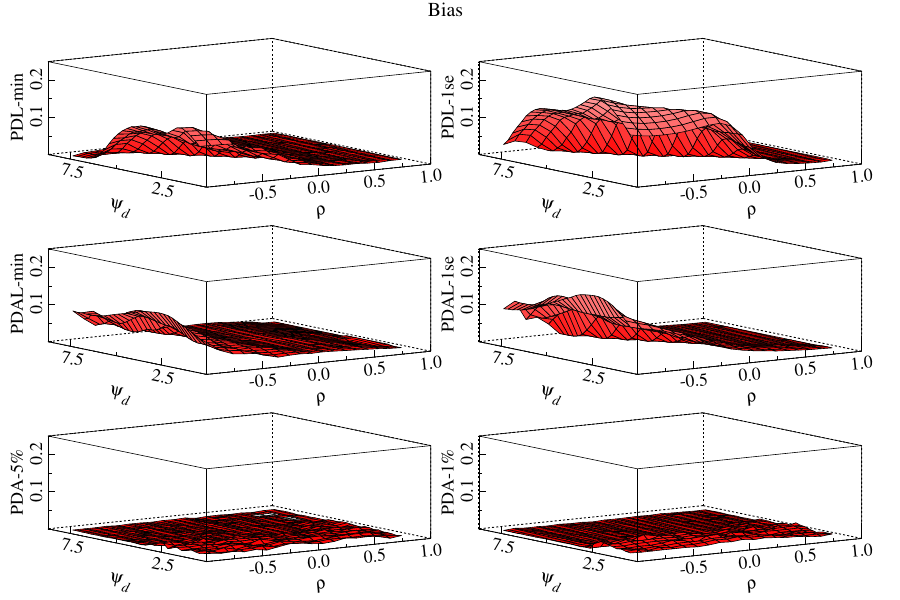} 
	\caption{Bias of $\hat\delta$ when $p>n$, with dependent covariates $\rho \in  [-0.9 ; 0.9]$ and varying non-centrality measure $\psi^d \in [1 ; 8]$. Design: $\psi^y = 2.5$, $n=200$, $p=210$.}
%	\caption{Bias of $\hat\delta$ when $p>n$, with varying non-centrality measure $\psi^d \in [1 ; 8]$ and dependent covariates $\rho \in  [-0.9 ; 0.9]$. Design: $\psi^y = 2.5$, $n=200$, $p=210$.}
%	\includegraphics[width=\linewidth, height=.4\textheight]{./Graphs/Bias_n_rep_1000_n_200_p_10_k_200_psiYD_0_psiDX_4_s2eps_1_s2eta_1.pdf} 
%	\caption{Bias of $\delta$, with varying non-centrality measure $\psi^y \in [1 ; 8]$ and dependent control variables $\rho \in  [-0.9 ; 0.9]$. Design: $\psi^d = 4$, $n=200$, $p=210$.}
\label{fig:simus_3D_bias_n<p} %Get_results_all_PsiYX_2
\end{figure}

\begin{figure}[htbp]
	\centering
	\includegraphics[width=\linewidth, height=.4\textheight]{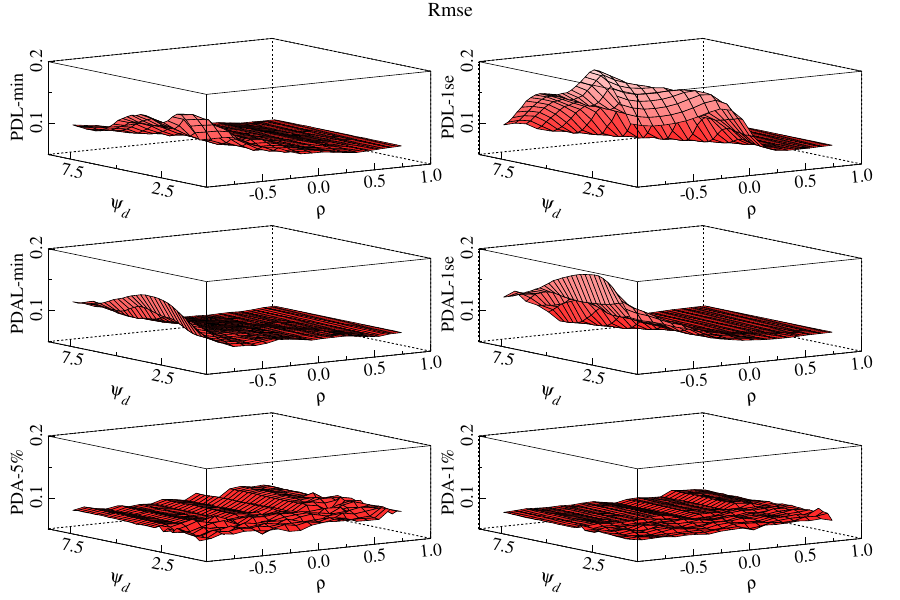} 
	\caption{RMSE of $\hat\delta$ when $p>n$, with dependent covariates $\rho \in  [-0.9 ; 0.9]$ and varying non-centrality measure $\psi^d \in [1 ; 8]$. Design: $\psi^y = 2.5$, $n=200$, $p=210$.}
%	\caption{RMSE of $\hat\delta$ when $p>n$, with varying non-centrality measure $\psi^d \in [1 ; 8]$ and dependent covariates $\rho \in  [-0.9 ; 0.9]$. Design: $\psi^y = 2.5$, $n=200$, $p=210$.}
%	\includegraphics[width=\linewidth, height=.4\textheight]{./Graphs/Rmse_n_rep_1000_n_200_p_10_k_200_psiYD_0_psiDX_4_s2eps_1_s2eta_1.pdf} 
%	\caption{RMSE of $\delta$, with varying non-centrality measure $\psi^y \in [1 ; 8]$ and dependent control variables $\rho \in  [-0.9 ; 0.9]$. Design: $\psi^d = 4$, $n=200$, $p=210$.}
	\label{fig:simus_3D_rmse_n<p} %Get_results_all_PsiYX_2
\end{figure}

\begin{figure}[htbp]
	\centering
	\includegraphics[width=\linewidth, height=.4\textheight]{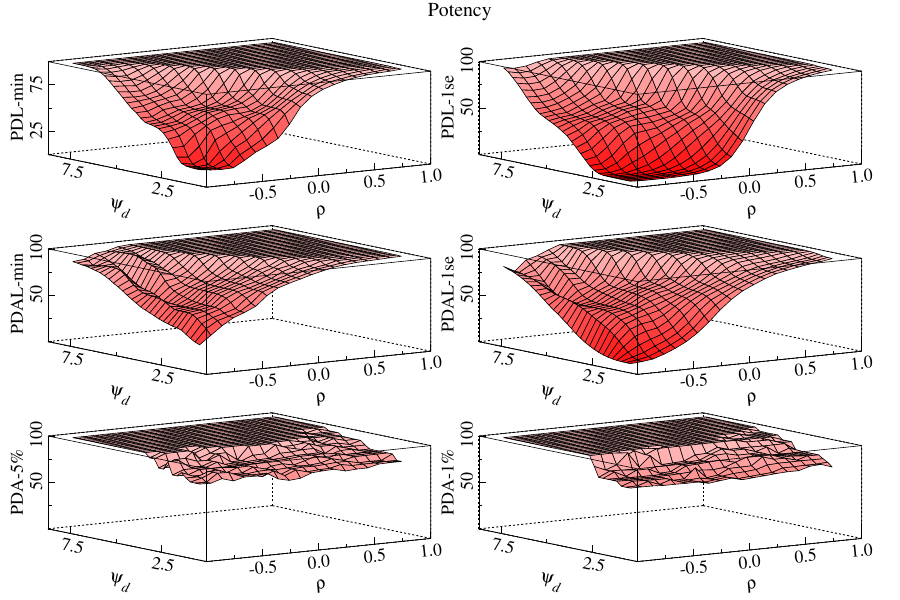} 
	\caption{Potency of $\hat\delta$ when $p>n$, with dependent covariates $\rho \in  [-0.9 ; 0.9]$ and varying non-centrality measure $\psi^d \in [1 ; 8]$. Design: $\psi^y = 2.5$, $n=200$, $p=210$.}
%	\caption{Potency of $\hat\delta$ when $p>n$, with varying non-centrality measure $\psi^d \in [1 ; 8]$ and dependent covariates $\rho \in  [-0.9 ; 0.9]$. Design: $\psi^y = 2.5$, $n=200$, $p=210$.}
%	\includegraphics[width=\linewidth, height=.4\textheight]{./Graphs/Potency_n_rep_1000_n_200_p_10_k_200_psiYD_0_psiDX_4_s2eps_1_s2eta_1.pdf} 
%	\caption{Potency of $\delta$, with varying non-centrality measure $\psi^y \in [1 ; 8]$ and dependent control variables $\rho \in  [-0.9 ; 0.9]$. Design: $\psi^d = 4$, $n=200$, $p=210$.}
	\label{fig:simus_3D_potency_n<p} %Get_results_all_PsiYX_2
\end{figure}

\begin{figure}[htbp]
	\centering
	\includegraphics[width=\linewidth, height=.4\textheight]{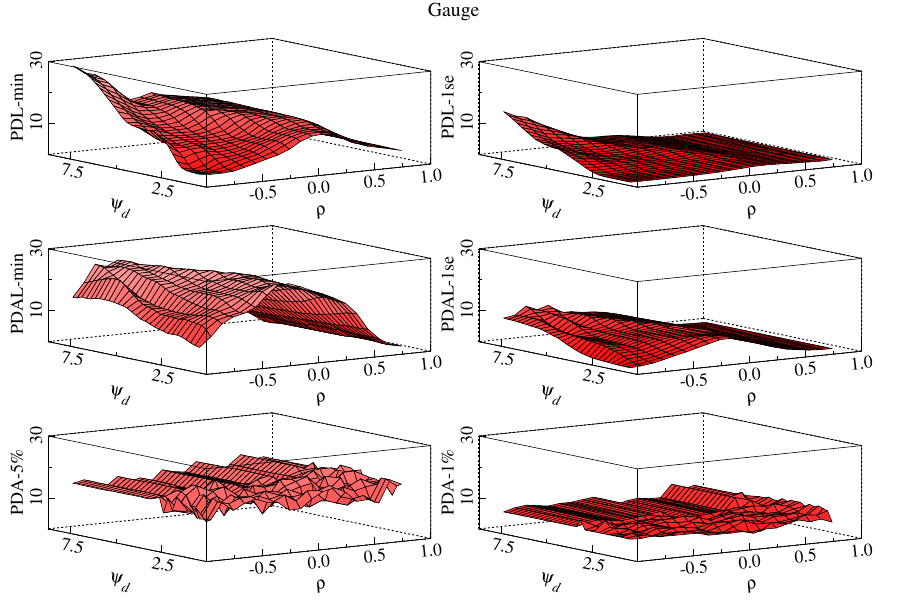} 
	\caption{Gauge of $\hat\delta$ when $p>n$, with dependent covariates $\rho \in  [-0.9 ; 0.9]$ and varying non-centrality measure $\psi^d \in [1 ; 8]$. Design: $\psi^y = 2.5$, $n=200$, $p=210$.}
%	\caption{Gauge of $\hat\delta$ when $p>n$, with varying non-centrality measure $\psi^d \in [1 ; 8]$ and dependent covariates $\rho \in  [-0.9 ; 0.9]$. Design: $\psi^y = 2.5$, $n=200$, $p=210$.}
%	\includegraphics[width=\linewidth, height=.4\textheight]{./Graphs/Gauge_n_rep_1000_n_200_p_10_k_200_psiYD_0_psiDX_4_s2eps_1_s2eta_1.pdf} 
%	\caption{Gauge of $\delta$, with varying non-centrality measure $\psi^y \in [1 ; 8]$ and dependent control variables $\rho \in  [-0.9 ; 0.9]$. Design: $\psi^d = 4$, $n=200$, $p=210$.}
\label{fig:simus_3D_gauge_n<p} %Get_results_all_PsiYX_2
\end{figure}

Finally, we study the robustness of double-selection methods in a high-dimensional linear regression model with more variables than observations. More precisely, we set $n = 200$ and keep all other parameters unchanged, so that $p=210>n$. Figures \ref{fig:simus_3D_bias_n<p}-\ref{fig:simus_3D_gauge_n<p} display respectively the bias, RMSE, potency and gauge as a function of the correlation coefficient, with $\rho\in [-0.9,0.9]$, and $\psi^d \in [1,8]$, for $\psi^y = 2.5$.
Results are similar to those of Figures \ref{fig:simus_3D_bias}-\ref{fig:simus_3D_gauge}, but some differences can be observed.
%Results are similar to those of Figures \ref{fig:simus_3D_bias}-\ref{fig:simus_3D_gauge}, and suggest that results from double-selection methods are robust. However, some differences can be observed for lasso- and autometrics-based methods.

First, the bias of Lasso-based methods is substantially larger than 0 for $p>n$ despite a larger potency. Indeed, as the sample size $n$ decreased compared to the previous simulation but the non-centrality parameters $\psi^y$ and $\psi^d$ remained constant, the non-zeros coefficients in $\bbeta$ and $\bgamma$ increased. Therefore, as Lasso selects variables based on the magnitude of the penalized coefficients (in absolute value), more relevant control variables are included in the terminal model, resulting in a larger potency. However, the effect of missed relevant control variables also increases (as $\beta$ is larger), leading to a larger bias.

%World coordinates
% Z1 0.05 Z2 0.15
%Y-axis: 0 8.5 2 2 1
%X-axis: -1 1 -0.9 0.4 0.2
%Z-axis: 0.05 0.15 0.05 0.05 0.025
% Copy World coordinates Z + axis, Labelling

Second, the effect of the adaptive-Lasso over the Lasso is unclear for $p>n$. While the bias and RMSE of Post-Double-Adaptive-Lasso and Post Double-Lasso are similar for $\psi^d \le 4$, they differ for larger values of $\psi^d$. Indeed, the bias and RMSE of Post-Double-Adaptive-Lasso increases with $\psi^d$, whereas they decrease towards 0 for Post-Double-Lasso when $\psi^d>0$ as for the case $p<n$.

Third, Post-Double-Autometrics is still unbiased whatever the values of $\rho$ and $\psi^d$. As Autometrics selects variables based on their significance, potency and bias remain similar. However, the gauge of Post-Double-Autometrics is slightly larger than $2\alpha$. Such result comes from the block-search performed by Autometrics when $p> n$, which leads to a small increase of the gauge.

To conclude, Post-Double-Autometrics also outperforms Post-Double(-Adaptive)-Lasso when $p>n$ and represents a reliable alternative when OLS is unfeasible.

\section{Empirical Application}

In this section, we revisit an international economic growth example. In the empirical growth literature, estimating the effect of an initial level of GDP per capita on the growth rates of GDP per capita is a key issue. Indeed, an important finding of the Solow-Swan-Ramsey growth model is the hypothesis of convergence, stating that poorer countries should catch up with richer ones over time as they grow faster. Following this hypothesis, the effect of the initial level of GDP on its growth rate, for a given country, should be negative.
To assess whether or not this assumption holds empirically, we use $n = 90$ complete observations extracted from the database of \citet{Barro1994}.\footnote{The data is available in the R library \texttt{hdm} of \citet{hdm} using the command \texttt{data("GrowthData")}.} The original dataset consists of a panel of 138 countries for the period of 1960 to 1985, and includes $p = 60$ control variables ($\bX$), such as variables measuring the education, trade openness or mortality rate. The dependent variable ($y$) corresponds to the national growth rates in GDP per capita
for the periods 1965-1975 and 1975-1985, and the variable of interest ($d$) is the real GDP per capita in 1965. See Table \ref{tab:description} in Appendix 1 for a complete description of the variables included in the dataset.

Many contributions in the literature investigated the convergence hypothesis. Using a simple bivariate regression model of growth rates on the initial level of GDP, \citet{Barro1995} showed that the convergence hypothesis is rejected. However, this estimate may be biased as the model does not take into account other countries' characteristics potentially affecting the growth rate, the initial level of GDP, or both. The literature thus focused on estimating this effect conditional on some characteristics (\citealp{Levine1992, Sala1997, Sala2004}). Given that the number of control variables is close to the number of observations, covariate selection is critical and the findings of these articles were severely criticized as they relied on ad-hoc procedures. More recently,  using the same sample of $90$ countries than in our application, \citet{Belloni2011b} and \citet{Belloni2013c} shed light on the convergence hypothesis using Lasso-based approaches. The authors find that a negative and statistically significant effect of the initial level of GDP on its growth rate, supporting the convergence hypothesis of the Solow-Swan-Ramsey growth model. 
We revisit this application using the different methods presented in Section \ref{Methods_section} to show whether or not our results are supporting those of the literature, and ultimately if the convergence hypothesis is verified.

\begin{table}[tbp]
		\caption{Estimates, robust standard errors, $90\%$ confidence intervals of the effect of the initial level of GDP $\delta$, and number of selected variables $k^*$ for the application on the growth data. Coefficients not significantly different from 0 at the 10\% nominal level are displayed in bold.}
	\centering
	\renewcommand{\arraystretch}{1.0}
	%	\begin{threeparttable}
		\centering
		\small
%		\begin{threeparttable}
		\begin{tabular}{lr@{}c@{}rr}
			\hline
			%			\multicolumn{1}{l}{Model} & \multicolumn{1}{l}{$\#$ of covariates} & \multicolumn{1}{l}{Point estimate} & \multicolumn{1}{l}{Standard error}  \\
			\multicolumn{1}{l}{Method} & \multicolumn{1}{c}{$\hat\delta$} & \multicolumn{1}{c}{Robust s.e.} & \multicolumn{1}{c}{$90\%$ CI} & \multicolumn{1}{c}{$k^*$ }    \\  %$\#$ of covariates } 
			\hline
			OLS without control variables & $\mathbf{0.0013}$ & $\mathbf{0.0053}$ & $\mathbf{[-0.0074;0.0101]}$ & $0$ \\
			OLS with all control variables & $\mathbf{-0.0094}$ & $\mathbf{0.0324}$ & $\mathbf{[-0.0627;0.0439]}$ & $60$ \\
			OLS, HCK standard errors & $\mathbf{-0.0094}$ & $\mathbf{0.0354}$ & $\mathbf{[-0.0676;0.0489]}$ & $60$ \\
			\hline
			Post-Lasso, $\lambda^{bya}$ & $\mathbf{0.0013}$ & $\mathbf{0.0053}$ & $\mathbf{[-0.0074;0.0101]}$ & $0$ \\
			Post-Lasso, $\lambda^{min}$ & $-0.0510$ & $0.0138$ & $[-0.0737;-0.0282]$ & $20$ \\
			Post-Lasso, $\lambda^{1se}$ & $-0.0347$ & $0.0119$ & $[-0.0543;-0.0151]$ & $10$ \\
			Post-Lasso, $\lambda^{bcch}$ & $-0.0174$ & $0.0093$ & $[-0.0328;-0.0020]$ & $7$ \\
			\hline
			Post-Adaptive-Lasso, $\lambda^{min}$ & $-0.0429$ & $0.0136$ & $[-0.0652;-0.0205]$ & $12$\\
			Post-Adaptive-Lasso, $\lambda^{1se}$ & $-0.0439$ & $0.0120$ & $[-0.0637;-0.0241]$ & $7$ \\
			\hline
			Post-Double-Lasso, $\lambda^{bya}$ & $-0.0419$ & $0.0167$ & $[-0.0694;-0.0144]$ & $7$ \\
			Post-Double-Lasso, $\lambda^{min}$ & $\mathbf{-0.0314}$ & $\mathbf{0.0207}$ & $\mathbf{[-0.0655;0.0027]} $& $23$ \\
			Post-Double-Lasso, $\lambda^{1se}$ & $-0.0563$ & $0.0191$ & $[-0.0876;-0.0249]$ & $13$ \\
			Post-Double-Lasso, $\lambda^{bcch}$ & $-0.0500$ & $0.0158$ & $[-0.0760;-0.0240]$ & $7$ \\
			\hline
			Post-Double-Adaptive-Lasso, $\lambda^{min}$ & $-0.0446$ & $0.0168$ & $[-0.0723;-0.0169]$ & $21$ \\
			Post-Double-Adaptive-Lasso, $\lambda^{1se}$ & $-0.0538$ & $0.0171$ & $[-0.0819;-0.0258]$ & $7$ \\
			\hline
			Autometrics, $\alpha = 0.05$ & $\mathbf{-0.0054}$ & $\mathbf{0.0118}$ & $\mathbf{[-0.0248;0.0140]}$ & $12$ \\
			Autometrics, $\alpha = 0.01$ & $-0.0430$ & $0.0106$ & $[-0.0606;-0.0255]$ & $3$ \\
			\hline
			Post-Double-Autometrics, $\alpha = 0.05$ & $\mathbf{-0.0257}$ & $\mathbf{0.0188}$ & $\mathbf{[-0.0566;0.0052]}$ & $14$ \\
			Post-Double-Autometrics, $\alpha = 0.01$ & $\mathbf{-0.0143}$ & $\mathbf{0.0196}$ & $\mathbf{[-0.0465;0.0179]}$ & $11$ \\
			\hline
		\end{tabular}
%		\begin{tablenotes}[para,flushleft]
%			\small
%			\noindent Note: The standard errors reported are robust standard errors.
%		\end{tablenotes}
%		\end{threeparttable}
		\label{tab:appligrowth}
	\end{table}

Table \ref{tab:appligrowth} displays the results obtained for Post-Lasso and Post-Double-Lasso with $\lambda = \{\lambda^{bya}, \lambda^{min}, \lambda^{1se}, \lambda^{bcch}\}$,  Post-Adaptive-Lasso and Post-Double-Adaptive-Lasso with $\lambda = \{\lambda^{min}, \lambda^{1se}\}$, Autometrics and Post-Double-Autometrics with $\alpha = \{0.05, 0.01\}$, as well as OLS regressions with and without control variables. We also report the results of an OLS regression involving all control variables with the correction of \citet{Cattaneo2018} to control for high-dimension and heteroscedasticity. 
The results of the OLS methods without control variables suggest that the convergence hypothesis is rejected as the coefficient is not significant at the 10\% nominal level, which is consistent with the finding of \citet{Barro1995}. 
This result should be taken with caution as this model is subject to omitted variable bias due to the non-inclusion of any control variables. The inclusion of 60 control variables in the model (with or without correction of the standard error of $\hat\delta$)
does not change the conclusion. However, this result should be interpreted with caution due to the high number of variables in the model, leading to a relatively large standard error. 

%Given that the latter two models have many covariates, the estimate is not precise, as shown by the relatively large standard errors and confidence intervals.  

%The results show that the coefficient $\hat \delta$ is always negative, except for the two methods that do not select any covariates, OLS without control variables and Post-Lasso with $\lambda^{bya}$. Consistent with the findings of our Monte Carlo simulations, results of the Post-Lasso, Post-Double-Lasso and Autometrics  are very sensitive to the choice of $\lambda$ and $\alpha$. 
To obtain more accurate estimates of $\delta$, we rely on the methods presented in Section~\ref{Methods_section} to select the relevant control variables.
The results show that the coefficient $\hat \delta$ is always negative, except for the two methods that do not select any covariates, i.e., OLS without control variables and Post-Lasso with $\lambda^{bya}$. Consistent with the findings of our Monte Carlo simulations, results of the Post-Lasso and Post-Double-Lasso are very sensitive to the choice of $\lambda$. 
While $\delta$ is significantly different from 0 at the 10\% nominal level for both Post-Adaptive-Lasso and Post-Double-Adaptive-Lasso, the results for Post-Lasso and Post-Double-Lasso methods depend on the choice of $\lambda$. Specifically, $\delta$ is significantly different from 0 for all Post-Lasso and Post-Double-Lasso methods except for Post-Lasso with $\lambda^{bya}$ and Post-Double-Lasso with $\lambda^{min}$.

Among the Post-Double-Lasso methods, $\lambda^{min}$ selects more than 10 additional control variables compared 
with $\lambda^{bya}$, $\lambda^{1se}$ and $\lambda^{bcch}$ and out of these four methods is the only one allowing to reject the hypothesis of convergence (or equivalently to fail to reject the null hypothesis $H_0: \delta=0$). Naturally, decreasing the penalization parameter increases the number of selected control variables, which affects the statistical inference on $\delta$.\footnote{Similarly, \citet{Belloni2011b} show in their application that despite the fact that the number of selected control variables increases as $\lambda$  decreases, their main conclusion remained unaffected by the choice of $\lambda$.} 
It echoes with the recommendation of \citet{wuthrich2023} to check the sensibility of the results to the choice of $\lambda$ by varying $\lambda$ in a grid around the chosen `optimal' value. Similar results are observed for the Post-Double-Adaptive-Lasso, even though $\delta$  remains significant for both values of $\lambda$.%
\footnote{Additional results with Elastic Net are presented in Appendix 3. $\delta$ is significantly different from 0 for Post-ElasticNet, not for Post-Double-ElasticNet which gives the same results as OLS.}

Autometrics results differ between the two target sizes considered in the application. Specifically, while $\delta$ is  significantly different from 0 for $\alpha=0.01$ (with only 3 control variables selected), $\delta$ is not significantly different from 0 for $\alpha=0.05$ (with 12 control variables selected).
 
%Furthermore, the results of the Post-Double-Lasso is also sensitive to $\lambda$. Indeed, while the Post-Double-Lasso with $\lambda^{bya}$, $\lambda^{1se}$ and $\lambda^{bcch}$ and Post-Double-Adaptive-Lasso with $\lambda^{min}$ and $\lambda^{1se}$ provide negative and significant coefficients, the Post-Double-Lasso leads to an insignificant coefficient with $\lambda^{min}$. 
%The conclusion regarding the convergence hypothesis thus depends on the choice of $\lambda$, which is a major issue as the optimal choice of $\lambda$ is unknown in practice. 

In contrast, Post-Double-Autometrics provides consistent results for both values of $\alpha$ and suggests that the convergence hypothesis is not verified (i.e., $H_0:\delta=0$ is not rejected). This conclusion does not support the previous findings of the literature obtained with Lasso-based methods.\footnote{Note that our results are also consistent with those of the Post-Double-Lasso with $\lambda^{min}$, which has been shown to be the overall best Lasso-based method in our previous section across several Monte Carlo simulation experiments.} Moreover, the results obtained are more accurate than those of OLS methods with all control variables as the standard errors of $\hat\delta$ are much smaller.

%REWRITE:
%The number of control variables selected by the Post-Double-Lasso also provides some information regarding its inconsistency. 

%%%%%%%%%%%%%%%%%%%%%%%%%%%%%%%%%%%%%%%%%%%%%%%%%%%%%%%%%%%%%%%%%%%%%%%%%%%%%%%
%%%%%%%%%%%%%%%%%%%%%%% Elastic-net based methods %%%%%%%%%%%%%%%%%%%%%%%%%%%%
%%%%%%%%%%%%%%%%%%%%%%%%%%%%%%%%%%%%%%%%%%%%%%%%%%%%%%%%%%%%%%%%%%%%%%%%%%%%%%%
% We also display, Table \ref{tab:appligrowth_ENET} in Appendix \ref{Appendix:add_tab_fig_emp}, the results obtained for Post-Elastic-Net and Post-Double-Elastic-Net methods. 
%%%%%%%%%%%%%%%%%%%%%%%%%%%%%%%%%%%%%%%%%%%%%%%%%%%%%%%%%%%%%%%%%%%%%%%%%%%%%%%
%%%%%%%%%%%%%%%%%%%%%%%%%%%%%%%%%%%%%%%%%%%%%%%%%%%%%%%%%%%%%%%%%%%%%%%%%%%%%%%
%%%%%%%%%%%%%%%%%%%%%%%%%%%%%%%%%%%%%%%%%%%%%%%%%%%%%%%%%%%%%%%%%%%%%%%%%%%%%%%

\begin{figure}
	\centering
\resizebox{15cm}{!}{	
	\includegraphics[width=\linewidth, height=.4\textheight]{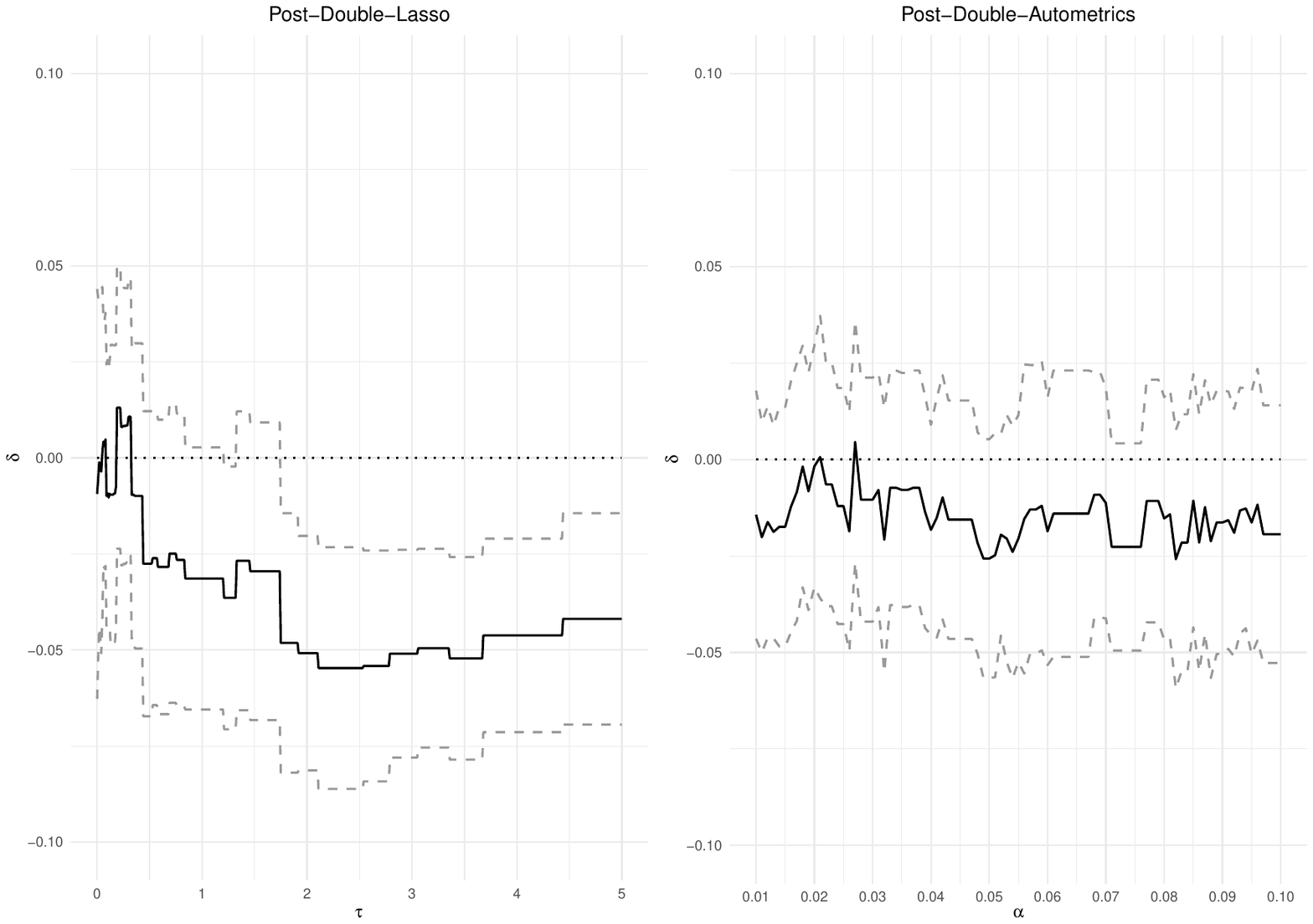}
}
	\caption{Estimate (solid line) and $90\%$ confidence interval (dashed line) of $\delta$ with the Post-Double-Lasso and penalty term $\lambda=\tau\lambda^{min}$ for different values of $\tau$ (Left), and with Post-Double-Autometrics for different values of $\alpha$ (Right).}
	\label{fig:appli}
\end{figure}

To illustrate the (lack of) robustness of the Post-Double-Lasso, Figure \ref{fig:appli} (left panel) displays the estimates and $90\%$ confidence intervals of $\hat\delta$ with $\lambda$ in the first two steps of the Post-Double-Lasso set to $\tau\lambda^{min}$ with $\tau \in \left(0,5\right)$.\footnote{Note that when $\tau = 0$, the results are identical to those of the OLS method with all control variables, and when $\tau = 1$ the results are those displayed in Table \ref{tab:appligrowth}.} The results show the sensitivity of the statistical inference on $\delta$ for Post-Double-Lasso to the value of $\lambda$. Indeed, while the coefficient is not significant for $\tau < 1.75$, it becomes negative and significant for $\tau \ge 1.75$.  
Results obtained for $\lambda^{bya}$, $\lambda^{1se}$, $\lambda^{bcch}$ are qualitatively similar (not reported here) %and available upon request.
%Qualitatively similar results are observed for $\lambda^{bya}$, $\lambda^{1se}$, $\lambda^{bcch}$ and for the Post-Double-Adaptive-Lasso with $\lambda^{min}$ and $\lambda^{1se}$ in Figures \ref{fig:lambdabyatheta}-\ref{fig:lambda1sealtheta} in Appendix \ref{Appendix:add_tab_fig_emp}.
Therefore, the conclusion of Post-Double-Lasso and Post-Double-Adaptive-Lasso regarding the convergence hypothesis strongly depends on the choice of $\lambda$. 

On the other hand, the results of Post-Double-Autometrics are robust to the choice of $\alpha$. Figure \ref{fig:appli} (right panel) displays the estimate and $90\%$ confidence interval of $\hat\delta$ for $\alpha \in [1\%,10\%]$. The results show that the estimates and confidence intervals are stable across all considered values of $\alpha$, and that the convergence hypothesis is never satisfied. In summary, unlike \citet{Belloni2011b} and \citet{Belloni2013c}, our results do not support the convergence hypothesis of the Solow-Swan-Ramsey growth model.%, and we recommend to rely on Post-Double-Autometrics rather than Post-Double-Lasso for its robustness. 

\begin{table}[tbp]
	\scriptsize
	\centering
%	\footnotesize
%	\renewcommand{\arraystretch}{1.0}
	\caption{Variables selected by Post-Double-Autometrics, Post-Double Lasso and Post-Double-Adaptive-Lasso methods.}
	\begin{threeparttable}
\begin{adjustbox}{width=.6\textwidth,center}
%\resizebox{9cm}{!}{
 		\begin{tabular}{lcccccccc}
		\hline 
		& \multicolumn{2}{c}{PDA} & \multicolumn{4}{c}{PDL} & \multicolumn{2}{c}{PDAL} \\ 
		& \multicolumn{1}{l}{$\alpha = 0.05$} & \multicolumn{1}{l}{$\alpha = 0.01$} & \multicolumn{1}{l}{$\lambda^{bya}$} & \multicolumn{1}{l}{$\lambda^{min}$} & \multicolumn{1}{l}{$\lambda^{1se}$} & \multicolumn{1}{l}{$\lambda^{bcch}$} & \multicolumn{1}{l}{$\lambda^{min}$} & \multicolumn{1}{l}{$\lambda^{1se}$} \\ 
		\hline
		bmp1l & \checkmark & \checkmark &       & \checkmark & \checkmark & \checkmark & \checkmark     &  \\
		pf65  & \checkmark & \checkmark &       &       &       &       &       &  \\
		lifee065 & \checkmark & \checkmark & \checkmark & \checkmark & \checkmark & \checkmark & \checkmark     & \checkmark \\
		mort1 & \checkmark & \checkmark &       &       &       &       & \checkmark     &  \\
		worker65 & \checkmark & \checkmark &       & \checkmark & \checkmark &       & \checkmark     & \checkmark \\
		ex1   & \checkmark & \checkmark &       & \checkmark & \checkmark &       & \checkmark     &  \\
		im1   & \checkmark & \checkmark &       & \checkmark &       &       &       &  \\
		mort65 & \checkmark &       &       &       &       &       &       & \checkmark \\
		gpop1 & \checkmark &       &       &       &       &       & \checkmark     &  \\
		no65  & \checkmark &       &       &       &       &       &       &  \\
		nof65 & \checkmark &       &       & \checkmark & \checkmark &       &       &  \\
		pop1565 & \checkmark &       &       & \checkmark &       &       & \checkmark    &  \\
		secc65 & \checkmark &       &       &       &       &       &       &  \\
		seccm65 & \checkmark &       &       & \checkmark &       &       &       &  \\
		sf65  &       & \checkmark & \checkmark & \checkmark &       & \checkmark &       &  \\
		pop6565 &       & \checkmark & \checkmark & \checkmark & \checkmark & \checkmark & \checkmark    & \checkmark \\
		syr65 &       & \checkmark &       &       &       &       &       &  \\
		syrf65 &       & \checkmark &       &       &       &       &       &  \\
		freeop &       &       &       &       &       &       & \checkmark     &  \\
		freetar &       &       & \checkmark & \checkmark & \checkmark & \checkmark & \checkmark     & \checkmark \\
		h65   &       &       &       &       &       &       & \checkmark     &  \\
		hm65  &       &       & \checkmark & \checkmark & \checkmark & \checkmark & \checkmark     & \checkmark \\
		hf65  &       &       &       & \checkmark &       &       & \checkmark     &  \\
		pm65  &       &       &       & \checkmark &       &       & \checkmark     &  \\
		geetot1 &       &       &       &       &       &       & \checkmark     &  \\
		geerec1 &       &       &       & \checkmark &       &       & \checkmark     &  \\
		gde1  &       &       &       & \checkmark &       &       & \checkmark     &  \\
		govwb1 &       &       &       &       &       &       & \checkmark     &  \\
		govsh41 &       &       &       & \checkmark & \checkmark &       &       & \checkmark \\
		gvxdxe41 &       &       &       &       & \checkmark &       & \checkmark     &  \\
		humanf65 &       &       & \checkmark & \checkmark & \checkmark & \checkmark &       &  \\
		pinstab1 &       &       &       & \checkmark &       &       & \checkmark     &  \\
		seccf65 &       &       &       & \checkmark &       &       &       &  \\
		teapri65 &       &       & \checkmark & \checkmark & \checkmark &       &       &  \\
		teasec65 &       &       &       & \checkmark & \checkmark &       &       &  \\
		xr65  &       &       &       & \checkmark &       &       &       &  \\
		tot1  &       &       &       &       &       &       & \checkmark     &  \\
			\hline 
       \multicolumn{9}{l}{Note: PDA, PDL and PDAL refer to Post-Double-Autometrics, Post-Double-Lasso,}\\
       \multicolumn{9}{l}{and Post-Double-Adaptive-Lasso methods, respectively. The symbol ``\checkmark'' indicates}\\
       \multicolumn{9}{l}{that a variable is selected by the corresponding method. }\\

		\end{tabular}
%}		
\end{adjustbox}
%       \smallskip\footnotesize\centering % <-- note the '\centering' directive
%        \smallskip         \begin{tablenotes}
%        \footnotesize \hspace{2cm}
%
%        \item[a] Note: PDA, PDL and PDAL refer to Post-Double-Autometrics, Post-Double-Lasso and Post-Double-Adaptive-Lasso methods, respectively. The symbol ``\checkmark'' indicates that a variable is selected by the corresponding method.
%        \end{tablenotes}

%		\begin{tablenotes}[para,center]
%		\small
%		\noindent Note: PDA, PDL and PDAL refer to Post-Double-Autometrics, Post-Double-Lasso and Post-Double-Adaptive-Lasso methods, respectively. The symbol ``\checkmark'' indicates that a variable is selected by the corresponding method.
%		\end{tablenotes}
	\end{threeparttable}

	\label{tab:va}
\end{table}

Finally, Table \ref{tab:va} displays the control variables selected by Post-Double-Autometrics, Post-Double-Lasso and  Post-Double-Adaptive-Lasso. Among all control variables, the most important ones are the black market premium (\textit{bmp1l}), which is a proxy of trade openness, the life expectancy at birth (\textit{lifee065}), the percentage of workers within the population (\textit{worker65}), and the ratio of export to GDP (\textit{ex1}) as they are selected by several methods. 
Interestingly, the variable measuring the ratio of import to GDP (\textit{im1}) is only selected by the three double-methods for which the convergence hypothesis is not verified, i.e., Post-Double-Autometrics with $\alpha = \{0.05, 0.01\}$ and Post-Double-Lasso with $\lambda^{min}$. 
This variable is likely the primary factor influencing the significance or non-significance of $\delta$ in the competing post-double methods, thereby contributing to the OVB. %This variable is most likely the source of the significance or non-significance of $\delta$ in the competing post-double methods, and is thus responsible for the omitted variable bias.
To validate this finding, we analyze the role of variable \textit{im1} in the model.

First, Table \ref{tab:t_stat_im1} reports the $t$-statistic associated with the coefficient of \textit{im1} in the two steps of Post-Double-Autometrics with $\alpha = \{0.05, 0.01\}$. The results confirm that \textit{im1} is relevant, as it is selected in both steps. While the $t$-statistics are positive in the first step, they are negative in the second step. This suggests that, conditional on the other selected control variables, the correlation between $d$ and \textit{im1} is negative. Furthermore, the $t$-statistics are relatively small (ranging from 2 to 4 in absolute value). This mirrors previous Monte Carlo simulation findings, where we found that when the non-centrality parameters of relevant variables were small  both in Equations \eqref{eqY} and \eqref{eqD}, Post-Double-Autometrics selects relevant variables more frequently than Post-Double-Lasso.

Second, Table \ref{tab:CI_with_without_im1} shows the confidence intervals for the three methods, with and without variable \textit{im1} in the final OLS regression. Interestingly, while the estimated effect of initial GDP, i.e., $\delta$, is not significant when \textit{im1} is included in the final OLS regression, it becomes significant when it is omitted. $\hat \delta$ remains negative, regardless of its significance, but the coefficient of \textit{im1} is positive when included in the final model (0.274, 0.355, and 0.444 for Post-Double-Lasso with $\lambda^{min}$ and Post-Double-Autometrics at $\alpha = \{0.05, 0.01\}$, respectively), leading to an insignificant $\delta$. This suggests that omitting \textit{im1} from the model introduces a negative bias. This finding highlights the importance of including \textit{im1} and explains why models excluding this variable result in a significant coefficient for initial GDP. Contrary to \Citet{Levine1992}, we thus find that the share of imports is not substitutable to export share, but that each of these two variables should be included in the terminal model.

\begin{table}[tbp]
	\caption{$t$-statistics of the variable \textit{im1} in the first two steps of the Post-Double-Autometrics method.}
	\label{tab:t_stat_im1}
	\centering
	\begin{threeparttable}
		\begin{tabular}{lcc}
			\hline
			Model & PDA, $\alpha = 0.05$ & PDA, $\alpha = 0.01$ \\
			\hline
			Step 1 &  2.6338 & 2.1340 \\
			Step 2 & -2.5068 & -3.8457 \\
			\hline
		\end{tabular}
		\begin{tablenotes}[para,flushleft]
			\small
			%		\noindent Note: PDA refers to Post-Double-Autometrics.
		\end{tablenotes}
	\end{threeparttable}
\end{table}

\begin{table}[tbp]
	\caption{Confidence interval with and without the variable \textit{im1} included in the final OLS regression. Coefficients not significantly different from 0 at the 10\% nominal level are displayed in bold.}
	\centering
	\begin{threeparttable}
	\begin{tabular}{l@{  }ccc@{  }cc}
		\hline
		 &  \multicolumn{2}{c}{with \textit{im1}} && 	 \multicolumn{2}{c}{without \textit{im1}}  \\
	\cline{2-3} \cline{5-6} \\[-2ex]
		 & $\hat{\delta}$  & \multicolumn{1}{c}{CI} && 	$\hat{\delta}$  & \multicolumn{1}{c}{CI } \\
%		 & $\hat{\delta}$ with \textit{im1} & \multicolumn{1}{c}{CI with \textit{im1}} & 	$\hat{\delta}$ without \textit{im1} & \multicolumn{1}{c}{CI without \textit{im1}} \\
		\hline
		PDL, $\lambda^{min}$ & $\mathbf{-0.0314}$ & $\mathbf{[-0.0655;0.0027]}$ && -0.0458 & $[-0.0732;-0.0183]$ \\
		\multicolumn{1}{l}{PDA, $\alpha = 0.05$} & $\mathbf{-0.0257}$ & $\mathbf{[-0.0566;0.0052]}$ && -0.0413 & $[-0.0713; -0.0113]$ \\
		\multicolumn{1}{l}{PDA, $\alpha = 0.01$} & $\mathbf{-0.0143}$ & $\mathbf{[-0.0465;0.0179]}$ && -0.0393 & $[-0.0654;-0.0132]$ \\
		\hline
	\end{tabular}
	\begin{tablenotes}[para,flushleft]
		\small
%		\noindent Note: PDA refers to Post-Double-Autometrics.
	\end{tablenotes}
	\end{threeparttable}
	\label{tab:CI_with_without_im1}
\end{table}

Finally, these three methods also select other variables in common, such as variables related to the population not working (\textit{pop1565} and \textit{pop6565}) or to education (\textit{nof65}, \textit{seccm65} and \textit{sf65}). The results thus suggest that the most important covariates allowing to reject the convergence hypothesis are mainly related to trade openness, infant life expectancy, proportion of worker in the population, export and import (relative to the GDP), and education.\footnote{Note that one variable related to education (\textit{pf65}) is only selected by both Post-Double-Autometrics methods.} 
Furthermore, Post-Double-Lasso with $\lambda^{min}$  selects several additional variables compared to Post-Double-Autometrics. These ones are mainly related to education, but also to tariff restriction, political instablity, exchange rate and government expenditures. However, given that Post-Double-Lasso with $\lambda^{min}$ selects on average many irrelevant variables (see previous section), most of these variables are probably irrelevant and could be dropped from the analysis.

\section{Conclusion}

In this paper, we have investigated the shortcomings of the Post-Double-Lasso method for estimating a parameter of interest, as an average treatment effect, in linear regression models with numerous covariates. While Post-Double-Lasso has gained popularity for its ability to mitigate omitted variable bias and manage overfitting through Lasso variable selection, we highlighted its vulnerabilities, particularly its sensitivity to the choice of the regularization parameter and the possibility of substantial omitted variable bias when some of the variables are negatively correlated.

As an alternative to Post-Double-Lasso, we introduced Post-Double-Autometrics, which leverages the Autometrics variable selection method, a method based on statistical inference to select relevant variables. Our findings demonstrate that Post-Double-Autometrics can effectively mitigate bias in estimating treatment effects by retaining relevant variables that might be overlooked by Lasso-based methods. Through extensive Monte Carlo simulations and an empirical illustration, we showed that Post-Double-Autometrics outperforms Post-Double-Lasso, yielding more reliable estimates of the treatment effect.

\section*{Acknowledgments}

We would like to thank the participants of the São Paulo School of Advanced Science on High Dimensional Models (Brasil), the 2024 French Stata conference in applied econometrics (Marseille) and the 2024 QFFE conference (Marseille) for their valuable comments. %We also thank the research support of the French National Research Agency Grants ANR-21-CE26-0007-01 and ANR-17-EURE-0020.

\section*{Funding}

%We gratefully  acknowledge the research support of the French National Research Agency Grants ANR-21-CE26-0007-01 and ANR-17-EURE-0020.

The project leading to this publication has received funding from the French government under the “France 2030” investment plan managed by the French National Research Agency (reference :ANR-17-EURE-0020 and ANR-21-CE26-0007-01) and from Excellence Initiative of Aix-Marseille University - A*MIDEX.

\bibliographystyle{chicago}

\bibliography{Biblio_SH}

%\pagebreak

\section*{Appendix 1: Name and description of the variables of the application on growth data}\label{Appendix:name_vars}

\begin{table}[htbp]
	\centering
	\scriptsize
 	\caption{Name and description of the variables of the application on growth data}
\resizebox{14.4cm}{!}{
 	\begin{tabular}{ll}
		\hline
		Name & Description \\
		\hline
		 \multicolumn{2}{c}{\underline{Dependent variable}}\\
		 Outcome & National growth rates in GDP per capita for the periods 1965-1975 and 1975-1985 \\
	 	\multicolumn{2}{c}{\underline{Variable of interest}}\\
		 gdpsh465 & Real GDP per capita (1980 international prices) in 1965 \\
		\multicolumn{2}{c}{\underline{Control variables}}\\
		 bmp1l & Black market premium Log (1+BMP) \\
		 freeop & Measure of "Free trade openness \\
		 freetar & Measure of tariff restriction \\
		 h65   & Total gross enrollment ratio for higher education in 1965 \\
		 hm65  & Male gross enrollment ratio for higher education in 1965 \\
		 hf65  & Female gross enrollment ratio for higher education in 1965 \\
		 p65   & Total gross enrollment ratio for primary education in 1965 \\
		 pm65  & Male gross enrollment ratio for primary education in 1965 \\
		 pf65  & Female gross enrollment ratio for primary education in 1965 \\
		 s65   & Total gross enrollment ratio for secondary education in 1965 \\
		 sm65  & Male gross enrollment ratio for secondary education in 1965 \\
		 sf65  & Female gross enrollment ratio for secondary education in 1965 \\
		 fert65 & Total fertility rate (children per woman) in 1965 \\
		 mort65 & Infant Mortality Rate in 1965 \\
		 lifee065 & Life expectancy at age 0 in 1965 \\
		 gpop1 & Growth rate of population \\
		 fert1 & Total fertility rate (children per woman) \\
		 mort1 & Infant Mortality Rate (ages 0-1) \\
		 invsh41 & Ratio of real domestic investment (private plus public) to real GDP \\
		 geetot1 & Ratio of total nominal government expenditure on education to nominal GDP \\
		 geerec1 & Ratio of recurring nominal government expenditure on education to nominal GDP \\
		 gde1  & Ratio of nominal government expenditure on defense to nominal GDP \\
		 govwb1 & Ratio of nominal government "consumption" expenditure to nominal GDP (using current local currency) \\
		 govsh41 & Ratio of real government "consumption" expenditure to real GDP (Period average) \\
		 gvxdxe41 & Ratio of real government "consumption" expenditure net of spending on defense and on education to real GDP \\
		high65 & Percentage of "higher school attained" in the total pop in 1965 \\
		highm65 & Percentage of "higher school attained" in the male pop in 1965 \\
		highf65 & Percentage of "higher school attained" in the female pop in 1965 \\
		highc65 & Percentage of "higher school complete" in the total pop \\
		highcm65 & Percentage of "higher school complete" in the male pop \\
		highcf65 & Percentage of "higher school complete" in the female pop \\
		human65 & Average schooling years in the total population over age 25 in 1965 \\
		humanm65 & Average schooling years in the male population over age 25 in 1965 \\
		humanf65 & Average schooling years in the female population over age 25 in 1965 \\
		hyr65 & Average years of higher schooling in the total population over age 25 \\
		hyrm65 & Average years of higher schooling in the male population over age 25 \\
		hyrf65 & Average years of higher schooling in the female population over age 25 \\
		no65  & Percentage of "no schooling" in the total population \\
		nom65 & Percentage of "no schooling" in the male population \\
		nof65 & Percentage of "no schooling" in the female population \\
		pinstab1 & Measure of political instability \\
		pop65 & Total Population in 1965 \\
		worker65 & Ratio of total Workers to population \\
		pop1565 & Population Proportion under 15 in 1965 \\
		pop6565 & Population Proportion over 65 in 1965 \\
		sec65 & Percentage of "secondary school attained" in the total pop in 1965 \\
		secm65 & Percentage of "secondary school attained" in male  pop in 1965 \\
		secf65 & Percentage of "secondary school attained" in female pop in 1965 \\
		secc65 & Percentage of "secondary school complete" in the total pop in 1965 \\
		seccm65 & Percentage of "secondary school complete" in the male pop in 1965 \\
		seccf65 & Percentage of "secondary school complete" in female pop in 1965 \\
		syr65 & Average years of secondary schooling in the total population over age 25 in 1965 \\
		syrm65 & Average years of secondary schooling in the male population over age 25 in 1965 \\
		syrf65 & Average years of secondary schooling in the female population over age 25 in 1965 \\
		teapri65 & Pupil/Teacher Ratio in primary school \\
		teasec65 & Pupil/Teacher Ratio in secondary school \\
		ex1   & Ratio of export to GDP (in current international prices) \\
		 im1  & Ratio of import to GDP (in current international prices) \\
		xr65  & Exchange rate (domestic currency per U.S. dollar) in 1965 \\
		tot1  & Terms of trade shock (growth rate of export prices minus growth rate of import prices) \\
		\hline
	\end{tabular}
}
	\label{tab:description}
\end{table}

\newpage

\section*{Appendix 2: Additional simulation results with Elastic-net}\label{Appendix:add_simu}
\label{app:simu}

\newpage
\section*{Appendix 3: Additional results for the empirical illustration}\label{Appendix:add_tab_fig_emp}
%\section{Additional tables and figures for the empirical illustration}\label{Appendix:add_tab_fig_emp}
\begin{table}[h!]
	\caption{Estimates, robust standard errors, $90\%$ confidence intervals of the effect of the initial level of GDP $\delta$, and number of selected variables $k^*$ of elastic-net based methods for the application on the growth data. Coefficients not significantly different from 0 at the 10\% nominal level are displayed in bold.}
	\centering
	\renewcommand{\arraystretch}{1.0}
	%	\begin{threeparttable}
		\centering
		\small
		%		\begin{threeparttable}
			\begin{tabular}{lr@{}c@{}rr}
				\hline
				%			\multicolumn{1}{l}{Model} & \multicolumn{1}{l}{$\#$ of covariates} & \multicolumn{1}{l}{Point estimate} & \multicolumn{1}{l}{Standard error}  \\
				\multicolumn{1}{l}{Method} & \multicolumn{1}{c}{$\hat\delta$} & \multicolumn{1}{c}{Robust s.e.} & \multicolumn{1}{c}{$90\%$ CI} & \multicolumn{1}{c}{$k^*$ }    \\  %$\#$ of covariates } 
			\hline
			Post-Elastic-Net, $\lambda^{min}$ & $-0.0502$ & $0.0186$ & $[-0.0808;-0.0197]$ & $45$ \\
			Post-Elastic-Net, $\lambda^{1se}$ & $-0.0447$ & $0.0169$ & $[-0.0724;-0.0169]$ & $32$ \\
			\hline
			Post-Double-Elastic-Net, $\lambda^{min}$ & $\mathbf{-0.0094}$ & $\mathbf{0.0324}$ & $\mathbf{[-0.0627;0.0439]} $& $60$ \\
			Post-Double-Elastic-Net, $\lambda^{1se}$ & $\mathbf{-0.0094}$ & $\mathbf{0.0324}$ & $\mathbf{[-0.0627;0.0439]}$ & $60$ \\
			\hline
		\end{tabular}
		%		\begin{tablenotes}[para,flushleft]
			%			\small
			%			\noindent Note: The standard errors reported are robust standard errors.
			%		\end{tablenotes}
		%		\end{threeparttable}
	\label{tab:appligrowth_ENET}
\end{table}

\end{document}